%% file: Main.tex
\documentclass[a4paper]{article}

\usepackage[english]{babel}
\usepackage[utf8x]{inputenc}
\usepackage[T1]{fontenc}
\usepackage{fancyhdr}
\usepackage{adjustbox}

\usepackage[a4paper,top=3cm,bottom=2cm,left=3cm,right=3cm,marginparwidth=1.75cm]{geometry}

\usepackage{amsmath}
\usepackage{graphicx}
\usepackage{url}
\usepackage[colorinlistoftodos]{todonotes}
\usepackage[colorlinks=true, allcolors=blue]{hyperref}
\usepackage{authblk}
\usepackage{lineno}

\numberwithin{equation}{subsection}

\newcommand\snowmass{\begin{center}
\rule[-0.2in]{\hsize}{0.01in}\\
\vskip 0.1in 
Submitted to the Proceedings of \\ the U.S. Particle Physics Community Planning Exercise (Snowmass 2021)\\ 
\rule{\hsize}{0.01in}\\
\end{center}}

\newcommand\pubnumber{FERMILAB-FN-1155-AD-PPD-SQMS-TD}

\newcommand\pubblock{\rightline{
    \begin{tabular}{l} \pubnumber\\
    \end{tabular}}}

\begin{document}

\title{\textbf{Higgs-Energy LEptoN (HELEN) Collider based on advanced superconducting radio frequency technology}}

\author[1,2]{S.~Belomestnykh\footnote{corresponding author, email: sbelomes@fnal.gov}} 
\author[1]{P.C.~Bhat}
\author[1]{A.~Grassellino}
\author[1]{M.~Checchin} 
\author[3]{D.~Denisov}
\author[4]{R.L.~Geng}
\author[1]{S.~Jindariani} 
\author[5]{M.~Liepe} 
\author[1]{M.~Martinello} 
\author[1]{P.~Merkel} 
\author[1]{S.~Nagaitsev} 
\author[1,5]{H.~Padamsee} 
\author[1]{S.~Posen} 
\author[6]{R.A.~Rimmer}
\author[1]{A.~Romanenko}
\author[1]{V.~Shiltsev} 
\author[1]{A.~Valishev} 
\author[1]{V.~Yakovlev}

\affil[1]{Fermi National Accelerator Laboratory, Batavia, IL, USA}
\affil[2]{Stony Brook University, Stony Brook, NY, USA}
\affil[3]{Brookhaven National Laboratory, Upton, NY, USA}
\affil[4]{Oak Ridge National Laboratory, Oak Ridge, TN, USA}
\affil[5]{Cornell University, Ithaca, NY, USA}
\affil[6]{Thomas Jefferson National Accelerator Facility, Newport News, VA, USA}

\maketitle

\vspace{-12cm}
\pubblock

\vspace{12cm}
\snowmass

\begin{abstract}
This Snowmass 2021 contributed paper discusses a Higgs-Energy LEptoN (HELEN) $e^+e^-$ linear collider based on advances superconducting radio frequency technology. The proposed collider offers cost and AC power savings, smaller footprint (relative to the ILC), and could be built at Fermilab with an Interaction Region within the site boundaries. After the initial physics run at 250~GeV, the collider could be upgraded either to higher luminosity or to higher (up to 500~GeV) energies. If the ILC could not be realized in Japan in a timely fashion, the HELEN collider would be a viable option to build a Higgs factory in the U.S.
\end{abstract}

\newpage

\tableofcontents


\newpage

\section{Executive summary}
\label{sec:ES}

Since the discovery of the Higgs boson, there is great interest in the world collider physics community for an $e^+e^-$ collider operating at $\sqrt{s}=250$~GeV (Higgs factory) and above, to make precision measurements of the Higgs boson couplings and searches for new physics beyond the Standard Model.  The International Linear Collider (ILC) based on superconducting radio frequency (SRF) accelerator technology was proposed to be hosted in Japan, shortly after the discovery of the Higgs boson in 2012. However, efforts to launch the ILC project in Japan haven't yet been successful. 

In the meantime, there continues to be steady progress in developing the SRF technology with accelerating gradients demonstrated up to 50~MV/m while the ILC design requires 31.5~MV/m.  It is anticipated that with an aggressive R\&D program on traveling wave SRF structures and innovations in cavity surface treatments and processing, an accelerating gradient of about 70~MV/m can possibly be reached within the next 2--3 years. Further research in developing new SRF materials, specifically Nb$_3$Sn,  could enable ultimate gradients of 90~MV/m and higher. 

Anticipating these advances in the near future, we propose an SRF-based 250~GeV $e^+e^-$ linear collider (HELEN collider) that can be sited at Fermilab.  With the use of existing infrastructure and facilities at Fermilab, and the factor of two higher gradients, there could be significant cost reduction for the main linac relative to the ILC main linac cost. If this proposal is supported by Snowmass and P5, a Conceptual Design Report can be produced in just a couple of years and Technical Design Report in $\sim 5$~years. If the ILC project in Japan will not gain traction, the expertise accumulated by the world's ILC community -- in particular in the U.S. laboratories and universities (Cornell, Fermilab, JLAB, SLAC, ...) -- would allow rapid developing, prototyping, and testing of new SRF cavities and cryomodules. Fermilab has capabilities that support the full cycle of R\&D, production, and verification (including testing cryomodules with beam) at the SRF accelerator test facilities and FAST. If given high priority, the construction of the HELEN collider could start as early as 2031--2032 with first physics in $\sim 2040$.

The HELEN collider can be upgraded to higher luminosities in the same way as was proposed for the ILC or to higher energies either by extending the linacs or with higher accelerating gradients as they become available.

\color{red}

\color{black}

\section{Introduction}
\label{sec:Intro}
One of the top priorities for the international particle physics community is to make precision measurements of the Higgs boson properties and search for a new physics beyond the Standard Model using an $e^+e^-$ collider at the center-of-mass energy of 250~GeV (Higgs factory). For many years, since 2013, the International Linear Colloder (ILC) has been a forerunner proposal for such a machine. It's mature SRF technology has been ``shovel ready'' and indeed has been used to build such SRF linacs as European XFEL in Hamburg, Germany and LCLS-II at SLAC in the USA. Meanwhile, SRF R\&D teams around the world continued to make progress pushing the performance of SRF cavities to higher accelerating gradients and quality factors. 

In the present paper we propose to use recent and anticipated advances in the SRF technology for a more compact and cost-effective $e^+e^-$ linear collider which we named Higgs-Energy LEptoN (HELEN) collider. If the ILC can not be realized in Japan in a timely manner, the proposed here Higgs factory could be built after relatively short period of dedicated R\&D efforts. In this paper we briefly review the physics at Higgs factory. At the core of our proposal is the SRF technology. First, we discuss state-of-the-art and limitations of the SRF technology for ILC and then describe relevant advances in SRF beyond that. The HELEN collider is described in some details including tentative list of parameters, layout and possible siting, and potentials for luminosity and energy upgrades. Following that we discuss possible approaches to a detector for the collider. Section 7 is dedicated to an outline of the accelerator and detector R\&D objectives. Finally, we provide brief summary and conclusions.

\input{Physics}

\input{SRFTechnology}

\input{HELENcollider}

\input{Detector}

\input{RnD}

\section{Summary and conclusions}
In this Snowmass 2011 contributed paper we presented a proposal for a Higgs factory based on advanced SRF technology. Recent achievements and anticipated near-future progress in this technology allow us to consider a linear $e^+e^-$ collider with accelerating gradients in the range from 55 to 90~MV/m. We selected the SRF traveling wave structure as the baseline option. The HELEN collider resembles the ILC in many aspects and would offer a similar physics program. The benefits of the HELEN collider are twofold: i) it offers an estimated 26\% main linac cost saving with respect to the ILC main linac and ii) due to higher accelerating gradient, the collider's smaller footprint could potentially be implemented at Fermilab.

A dedicated R\&D efforts are still necessary to make further advances and demonstrate feasibility of the proposed approach. However, with an appropriate level of investment this focused program could be accomplished in $\sim~5$~years, after which a Technical Design report would be prepared. This effort is a part of the integrated U.S. collider R\&D program proposed in \cite{FutureColliderUS}.

\section{Acknowledgements}
Work supported by the Fermi National Accelerator Laboratory, managed and operated by Fermi Research Alliance, LLC under Contract No. DE-AC02-07CH11359 with the U.S. Department of Energy. The U.S. Government retains and the publisher, by accepting the article for publication, acknowledges that the U.S. Government retains a non-exclusive, paid-up, irrevocable, world-wide license to publish or reproduce the published form of this manuscript, or allow others to do so, for U.S. Government purposes.

\bibliographystyle{JHEP}
\bibliography{Bib.bib}

\end{document}

%% file: Physics.tex
\section{Physics at Higgs factory\label{sec:physics}}

Here we provide general qualitative statements about physics at the $e^+e^-$ Higgs factories. Detailed physics studies for the $e^+e^-$ colliders have been conducted by the ILC, FCC, CLIC, and CEPC collaborations~\cite{Baer:2013cma,FCC:2018byv,Linssen:2012hp,CEPC-SPPCStudyGroup:2015csa}. Physics reach of HELEN should be similar to ILC, assuming that the same integrated luminosity and beam polarization levels are achievable. 

The Standard Model (SM) of elementary particles has been validated extensively through precision experiments and found to be incredibly successful at describing our world. However, despite being internally consistent and very successful, there are a number of experimental observations that the SM fails to explain. It does not fully explain the baryon asymmetry, incorporate the theory of gravitation as described by general relativity, or account for the accelerating expansion of the Universe as possibly described by dark energy. The model does not contain any viable dark matter particle that possesses all of the required properties deduced from cosmology and astrophysics. It also does not incorporate neutrino oscillations and their non-zero masses. Furthermore, the model suffers from several internal shortcomings, such as the hierarchy problem, where fine-tuned cancellations of large quantum corrections are required in order for the Higgs boson mass to be near the electroweak scale. It is evident that the Standard Model is just an effective theory that appears, so far, to be valid at the energies experimentally accessible today.

For the next two decades, the LHC will remain the highest energy collider in the world. The full LHC dataset is expected to be 20 times more than what we have today.  Such a dataset will provide great opportunities for studies of the SM, including detailed characterization of the Higgs boson. Besides the precision, the LHC data will also greatly extend the sensitivity for new physics. However, it is conceivable that the HL-LHC dataset will not be sufficient to discover and fully characterize new physics. This provides a strong motivation for an $e^+e^-$ Higgs factory.

Detailed exploration of the electroweak sector of the Standard Model remains a high priority. An $e^+e^-$ Higgs factory will enable highly precise measurements of Standard Model parameters, which in turn provide deeper insight into the mechanism of electroweak symmetry breaking. This includes precise determination of the nature of the Higgs boson, including measurements of its properties and couplings. It has been demonstrated that a wide range of new physics models with multi-TeV scale result in few percent level modifications to the Higgs boson couplings~\cite{Dawson:2013bba}. Therefore, measuring Higgs boson couplings at the sub-percent level can provide first indirect evidence of beyond-SM (BSM) particles or forces. Measurements of the Higgs boson decay rate to invisible particles is also very important for discovering or constraining BSM physics. Beyond the couplings, measurement of the Higgs boson total width  and self-interactions (both trilinear and quartic) would further shed light on the underlying structure of the electroweak sector; these measurements (as well as top Yukawa coupling) however require collision energies of 500 GeV or higher.   

While the Higgs boson remains a centerpiece for the precision program at Higgs factories, many other rare SM processes continue to attract significant interest. For example, operation at lower energies on the $Z$ resonance (91 GeV) and at the $WW$ threshold (160 GeV) will allow to gather large amounts of data and perform precision measurements of the electroweak sector of the SM. An $e^+e^-$ collider with luminosity of $10^{34}$ cm$^{-2}$s$^{-1}$ will produce billions of $Z$ boson and tens of millions of $WW$ events. This dataset will allow to significantly improve current precision of the key electroweak observables, such as the weak mixing angle ($\sin^{2}(\theta_{eff})$), the masses and widths of the $W$ and $Z$ bosons, the forward-backward asymmetry for bottom quarks ($A_b$) and the polarization asymmetry of tau leptons ($A_\tau$). A very precise determination of the strong coupling constant $\alpha_s$ at the $M_Z$ scale and the number of weakly-interacting neutrinos are also possible.

Besides the precision, $e^+e^-$ colliders offer excellent opportunities for direct observation of new physics, covering many orders of magnitude of coupling strengths and mass scales. For example, signatures of dark photons and axion-like particles can be searched for in decays of the $Z$ bosons produced in the 91 GeV run. New resonances ($Z'$) with masses up to the collision energy and decaying into fermion ($f$) pairs, predicted by many extensions to the SM (e.g. compositeness, extra dimensions, etc), can be discovered in the $e^+e^{−} \rightarrow ff$ process. In SUSY, low momentum thresholds available at the $e^+e^-$ colliders enable excellent capability to look for naturally light and compressed electroweakino states that are very challenging at the hadron machines, thus providing a nice complementarity to the LHC searches. Finally, a wide class of BSM models with extended Higgs sectors can be probed by looking for pair production of the additional Higgs bosons. 

It is evident from the considerations above that HELEN collider opens window to a reach and exciting physics program, with excellent chances for fundamental discoveries.

%% file: SRFTechnology.tex
\section{SRF technology for linear colliders}
\label{sec:SRFtechnology}

\subsection{SRF for ILC: state of the art and limitations}

Superconducting Radio Frequency (SRF) technology for a linear collider has been in development since early 1990's \cite{1st_TESLA_ws}. The SRF option was selected for the International Linear Collider (ILC), which has been the prime candidate for a next lepton HEP collider, especially since the discovery of the Higgs boson in 2012. The machine was baselined in 2013 \cite{ILC_TDR-v3-I, ILC_TDR-v3-II} and is under consideration to be hosted in Japan. The collider facility will be about 20.5 kilometers in total length, and will accelerate beams of electrons and positrons to 125 GeV each to operate at the center-of-mass energy of 250 GeV, see e.g., \cite{ILC_Snowmass2021}. The design instantaneous luminosity of the collider will be $1.35\times10^{34}$ cm$^{-2}$s$^{-1}$ with proposals to upgrade to higher luminosity (up to $8.1\times10^{34}$ cm$^{-2}$s$^{-1}$ or, accounting for polarization, with an effective luminosity up to $2.0\times10^{35}$ cm$^{-2}$s$^{-1}$) \cite{padamsee2019impact}.

The baseline ILC SRF technology is well-established \cite{TESLAcavity} and was used to build such machines as European XFEL \cite{E-XFEL} and LCLS-II \cite{LCLS-II}. The ILC design specifies an accelerating gradient $E_{acc}$ of 31.5~MV/m and intrinsic cavity quality factor $Q$ of $1\times10^{10}$ per the ILC Technical Design Report (TDR) \cite{ILC_TDR-v3-I, ILC_TDR-v3-II}. During phase II of the ILC R\&D program, the $(94 \pm 6)$\% yield has been achieved for cavities that demonstrated accelerating gradients >28~MV/m and $(75\pm11$\% for 35~MV/m. These yields were reached after cavities with gradients outside the ILC specification have been re-treated. This ensemble of cavities has an average gradient of 37.1~MV/m. At DESY, two large-grain 9-cell cavities also reached 45~MV/m \cite{DESY_large_grain}.

\begin{figure}[tb]
    \centering
    \includegraphics[width=0.7\textwidth]{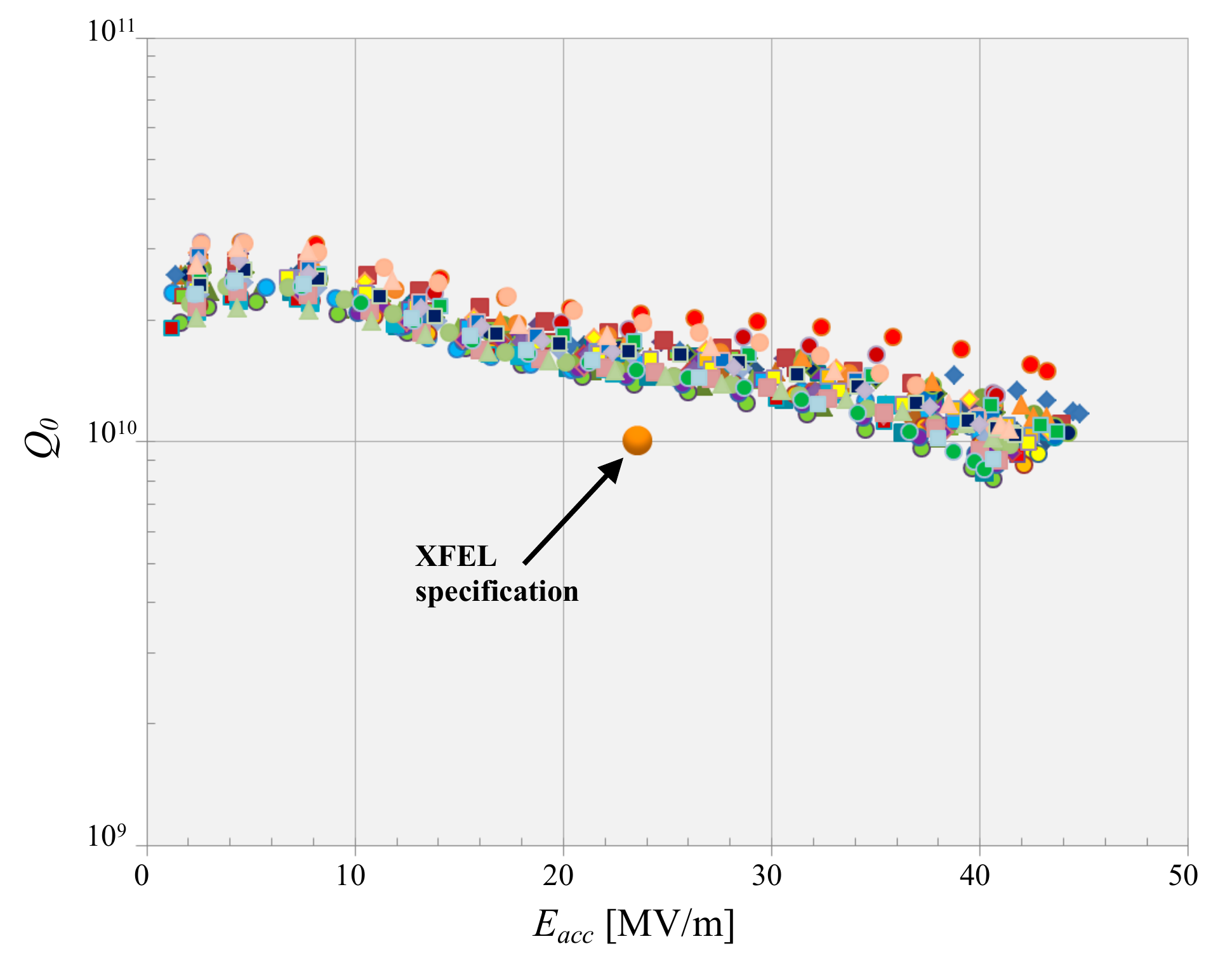}
    \caption{Performance of 47 best cavities from the European XFEL production run.}
    \label{fig:EXFEL_best_cavities}
\end{figure}

The cavity performance for the European XFEL cavities is close to the requirements for the ILC TDR.  
The 420 cavities from one vendor which followed the ILC EP recipe for final treatment succeeded in reaching an average gradient of $33.0~\pm$~6.5~MV/m. More than 10\% of cavities from this vendor exceeded 40~MV/m. Figure~\ref{fig:EXFEL_best_cavities} shows performance plots for 47 “best” cavities from the European XFEL production run with 40--45~MV/m \cite{best_EXFELcavities}. 
The average cryomodule gradient reached was $27.5~\pm~4.8$~MV/m. So, the cryomodule performance does need to be improved to reach ILC specs of 31.5~MV/m. It is very encouraging that 18 out of 97 cryomodules reached the operating gradient of >30~MV/m, which is close to the ILC-TDR specification. Bout 50\% of all cavities tested in 97 modules have operating gradients above or equal to the 31~MV/m administrative limit\cite{KasprzakSRF2017}.

The demonstrated success from the European XFEL cavity and cryomodule production is very encouraging for reaching ILC goals. It is rare that an approximately 10\% prototype demonstration exists for a new HEP machine, and for its core technology.

Two laboratories have continued efforts to demonstrate the ILC cryomodule (CM) goal, but on a smaller scale. An ILC demonstration cryomodule prepared by Fermilab reached the total available accelerating voltage of 267.9~MV, which is equivalent to an average gradient of 32.2~MV/m for eight 1.038~m ling cavities, thus exceeding the ILC goal. A beam with the 3.2~nC bunch charge has obtain an energy gain of >255~MeV after passing through the CM \cite{FASTrecord}. Nine (including one nitrogen-infused cavity) out of twelve cavities in a demonstration cryomodule at the KEK test facility (STF) exceeded the ILC specification, achieving an average accelerating gradient of 33~MV/m during beam operation \cite{SRF_2021}. Three of these cavities exceeded 36~MV/m. Cavities for a new high gradient demonstration module at Fermilab have reached 40 MV/m average in their vertical tests. The CM test has yet to be conducted.

The SRF technology for the ILC at 250~GeV is ``shovel-ready''. The technology has been demonstrated and industrialized. Besides European XFEL, new large scale facilities -- LCLS-II, ESS, PIP-II, and SHINE -- are soon to be commissioned or under construction. Extensive SRF infrastructure exists worldwide for cavity fabrication, surface treatment, clean assembly, cold testing, and cryomodule assembly. Major SRF facilities are available at DESY, CERN, INFN, CEA/Saclay, IJCLab/Orsay, KEK, JLAB, Cornell, Fermilab, MSU, and at several industries around the world. New infrastructure is becoming available for upcoming projects such as for ESS in Europe, and PAPS in China. New industries in South Korea, China, and Japan are rapidly growing familiar with SRF technology.

 With some modest investment, the ILC will be upgradeable to higher collision energies up to 380~GeV in the future \cite{padamsee2019impact}.  In principle, upgrades to 500~GeV, 1 TeV, and beyond are possible~\cite{ILC_Snowmass2021, padamsee2021ilc}.

\subsection{SRF technology advances beyond ILC}

\subsubsection{Nb-based standing wave SRF cavities}
\label{sec:SW_SRF}

Superconducting RF technology continues to move forward. In this section, we describe the on-going R\&D program to improve the gradient and $Q$ of standing wave (SW) SRF accelerating structures made of bulk niobium. The application of these newly developed techniques to travelling wave structures with lower peak fields than standing wave structures will open the door to 70~MV/m superconducting RF for HELEN, as will be described in section~\ref{sec:TW_SRF}.

\begin{figure}
    \centering
    \includegraphics[width=0.7\textwidth]{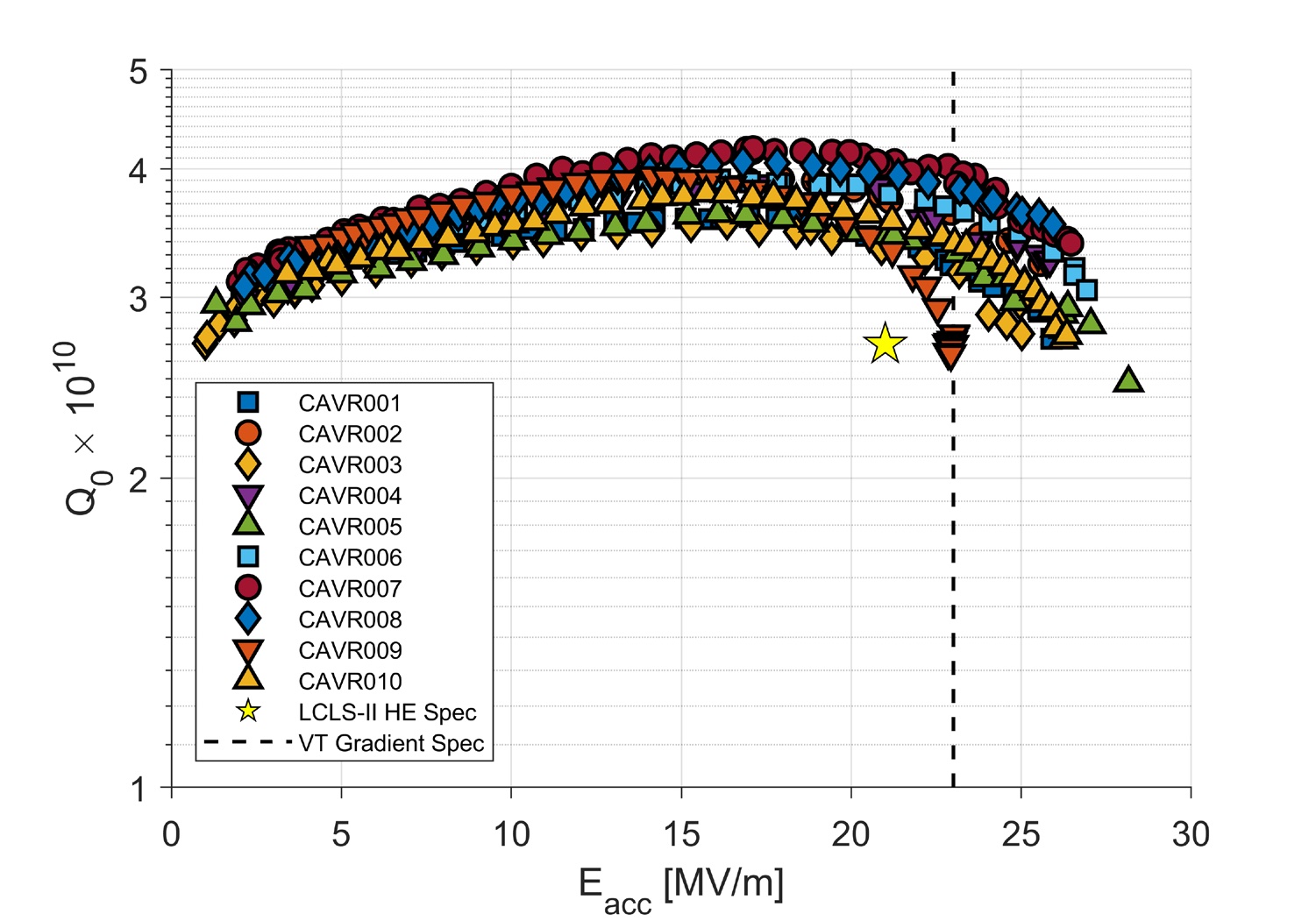}
    \caption{$Q$ vs. $E_{acc}$ curves for ten LCLS-II-HE 9-cell cavities processed with $2/0$ N-doping recipe followed by cold EP \cite{Gonnella_TTC2022}.}
    \label{fig:LCLS-II-HE_cavities}
\end{figure}

Key areas of further development over the last 5--10 years have been for higher $Q$ values at medium gradients (16--25~MV/m) for CW operation with the invention of new techniques of nitrogen doping (N-doping) \cite{N-doping}. A remarkable outcome of N-doping is the rise in $Q$ with field, as opposed to standard fall in $Q$ behavior with the ILC cavity surface treatment. Nitrogen doping for high $Q$ has already been applied to the construction of  LCLS-II, and its high energy upgrade LCLS-II-HE. For LCLS-II, more than 300 1.3-GHz cavities in 38 cryomodules, have been delivered to SLAC, and 35 are installed in the tunnel (the other 3 are spares). For LCLS-II-HE, ten 1.3~GHz 9-cell N-doped cavities have reached average gradient of 25.9~MV/m and average $Q$ of $3.6\times 10^{10}$ at 23~MV/m (the acceptance gradient for vertical cavity tests) (Figure~\ref{fig:LCLS-II-HE_cavities}) \cite{Gonnella_TTC2022}. 

\begin{figure}
    \centering
    \includegraphics[width=0.65\textwidth]{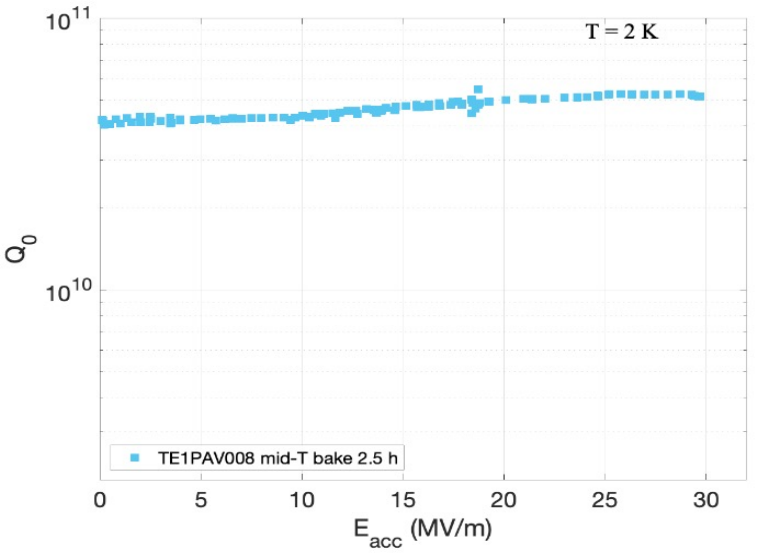}
    \caption{$Q = 5\times10^{10}$ at 30~MV/m by baking at $\sim300^{\circ}$C to dissolve the natural oxide (and other surface layers) into the bulk, but not exposing the cavity to air or water before RF measurements.}
    \label{fig:mid-T_baking}
\end{figure}

\begin{figure}
    \centering
    \includegraphics[width=0.7\textwidth]{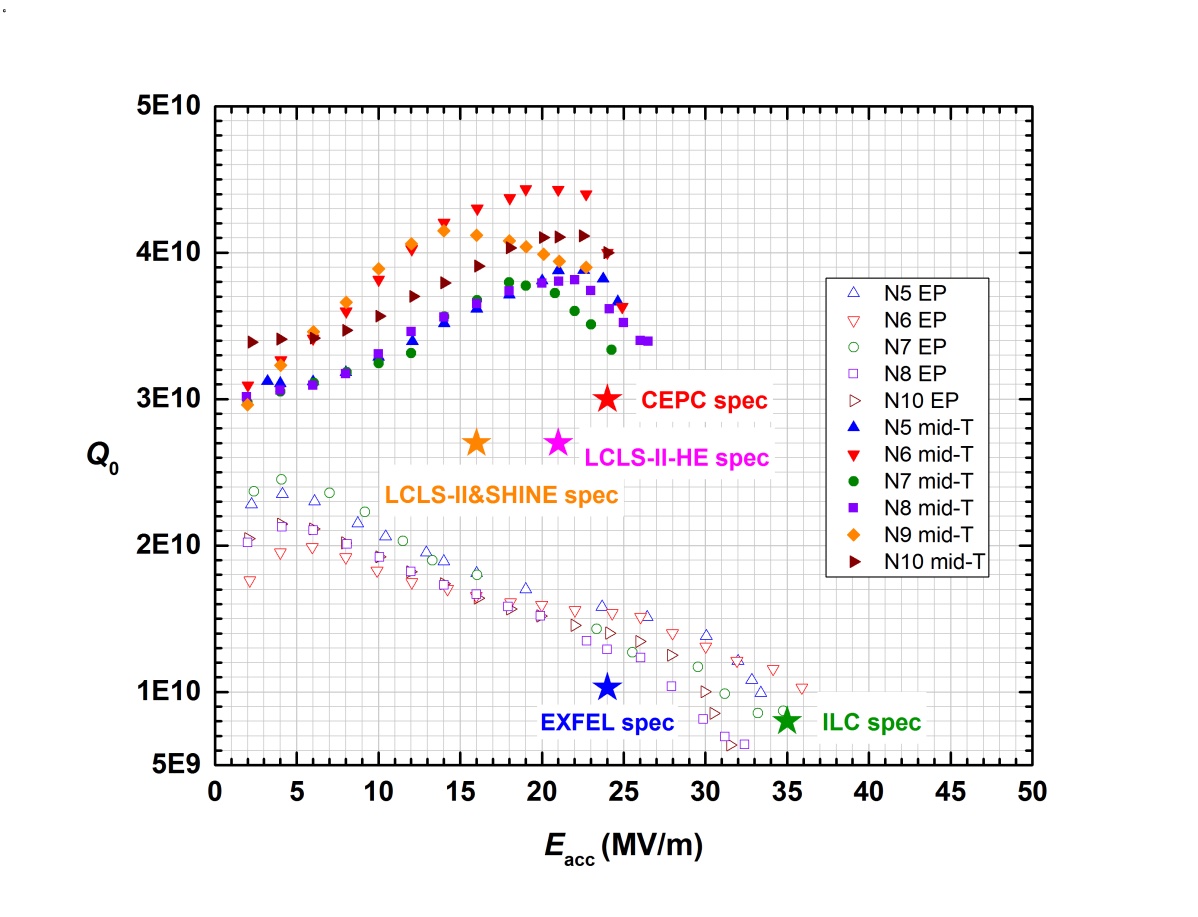}
    \caption{Results from IHEP (China) on mid--$T$ baking 9-cell cavities after exposure to air and high pressure water rinsing. The results are compared to those with the standard ILC treatment.}
    \label{fig:IHEP_mid-T_baking}
\end{figure}

Further $Q$ improvements come from exciting developments \cite{Posen_mid-T_baking} that show $Q = 5\times10^{10}$ at 30~MV/m by baking at $\sim300^{\circ}$C (mid-temperature, or mid--$T$, baking) to dissolve the natural oxide and other surface layers into the bulk. It is interesting to note (Figure~\ref{fig:mid-T_baking}) how the $Q$ rises with field, as seen for N-doping.  After exposure to air, followed by High-Pressure Rinsing with ultra-pure deionized water (HPR), the $Q$ dropped to $2\times10^{10}$ at 30~MV/m. Surface analysis of similarly treated samples show a nitrogen peak at a few nm below the surface, suggesting that nitrogen is naturally present at the surface and has diffused into the Nb to give the N-doping effect. IHEP in Beijing, China followed up on these encouraging results with several 9-cell TESLA cavities with similar exciting results \cite{he2020mediumtemperature}. After mid--$T$ ($300^{\circ}$C) furnace bake, and HPR, all the 9-cell cavities demonstrate high $Q$ in the range of $3.5 - 4.4\times10^{10}$ at the gradient between 16–-24~MV/m, as shown in Figure~\ref{fig:IHEP_mid-T_baking}. KEK is also pursuing the mid--$T$ baking option. Although in its early stages, the mid--$T$ baking procedure shows the potential of Nb for high gradients with high $Q$'s.

\begin{figure}
    \centering
    \includegraphics[width=\textwidth]{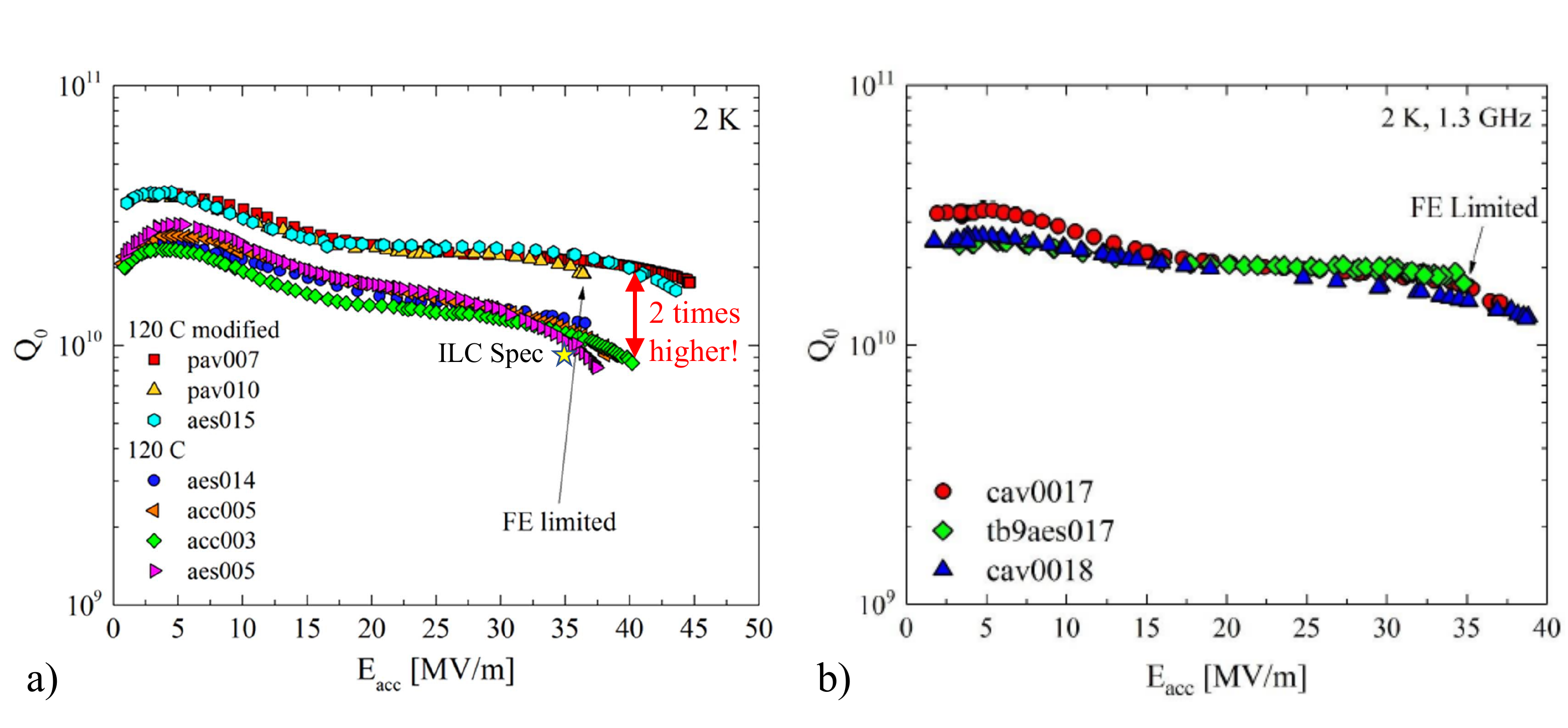}
    \caption{Nitrogen infusion results: (a) single-cell cavities, (b) 9-cell cavities.}
    \label{fig:Nitrogen_infusion}
\end{figure}

\begin{figure}
    \centering
    \includegraphics[width=\textwidth]{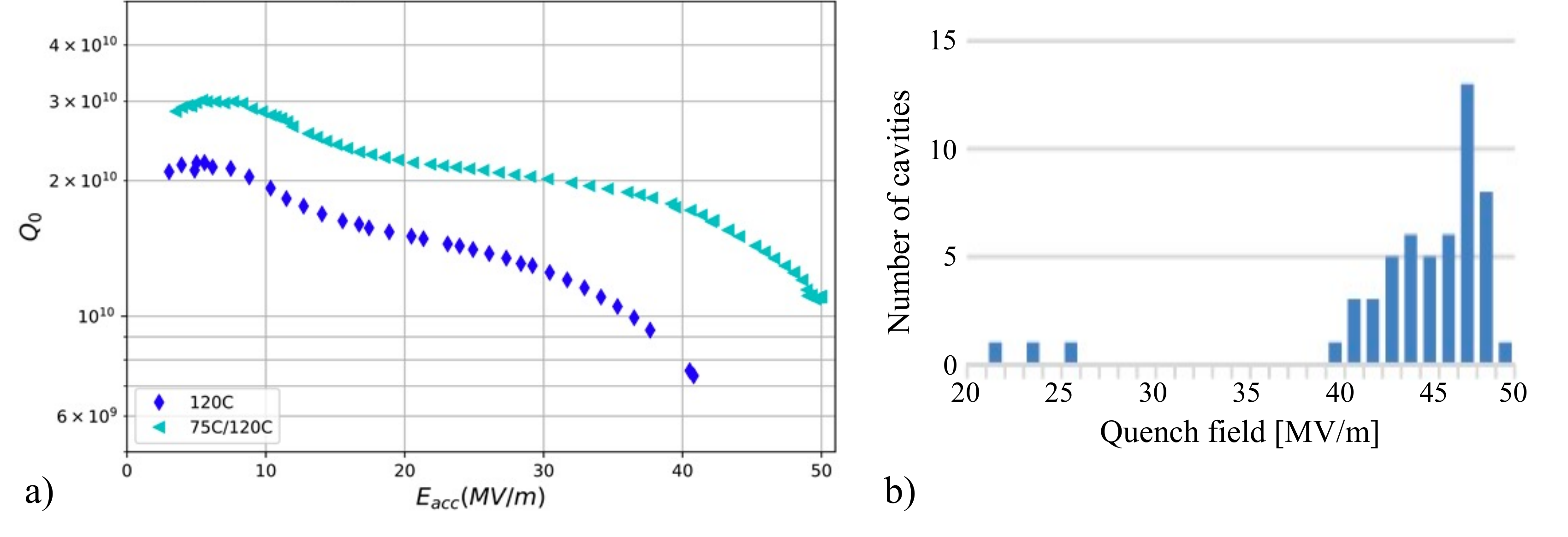}
    \caption{(a) $Q$~vs.~$E$ curve of single-cell cavity reaching 49~MV/m from cold EP/optimized baking ($75/120^{\circ}$C) compared to the curve of a cavity prepared by the standard ILC recipe. (b) Histogram of gradients of a large number of single-cell cavities prepared by cold EP/optimized baking ($75/120^{\circ}$C).}
    \label{fig:75/120C_bake}
\end{figure}

On the high gradient frontier (with higher $Q$’s), the invention of nitrogen infusion \cite{nitrogen_infusion} (N-infusion, stemming from the discovery of nitrogen doping) demonstrates gradients of 40–-45~MV/m as shown in Figure~\ref{fig:Nitrogen_infusion}, and compared to the lower performance of cavities prepared with the standard ILC recipe.
In yet another new and extraordinary development, quench fields near 50~MV/m for 1.3 GHz niobium TESLA-shaped SRF single-cell cavities have been achieved with a new $75/120^{\circ}$C two-step bake treatment developed at FNAL \cite{2-step_baking}, as shown in Figure~\ref{fig:75/120C_bake}(a). The two-step baking with cold electropolishing \cite{Bafia_SRF2019} shows gradients in the range of 40--50~MV/m (average 45~MV/m), as depicted by the histogram of about 50 tests in Figure~\ref{fig:75/120C_bake}(b). Note that 3 cavities that quench below 28~MV/m were found to have rare physical defects that likely limited their performance.

\begin{figure}
    \centering
    \includegraphics[width=\textwidth]{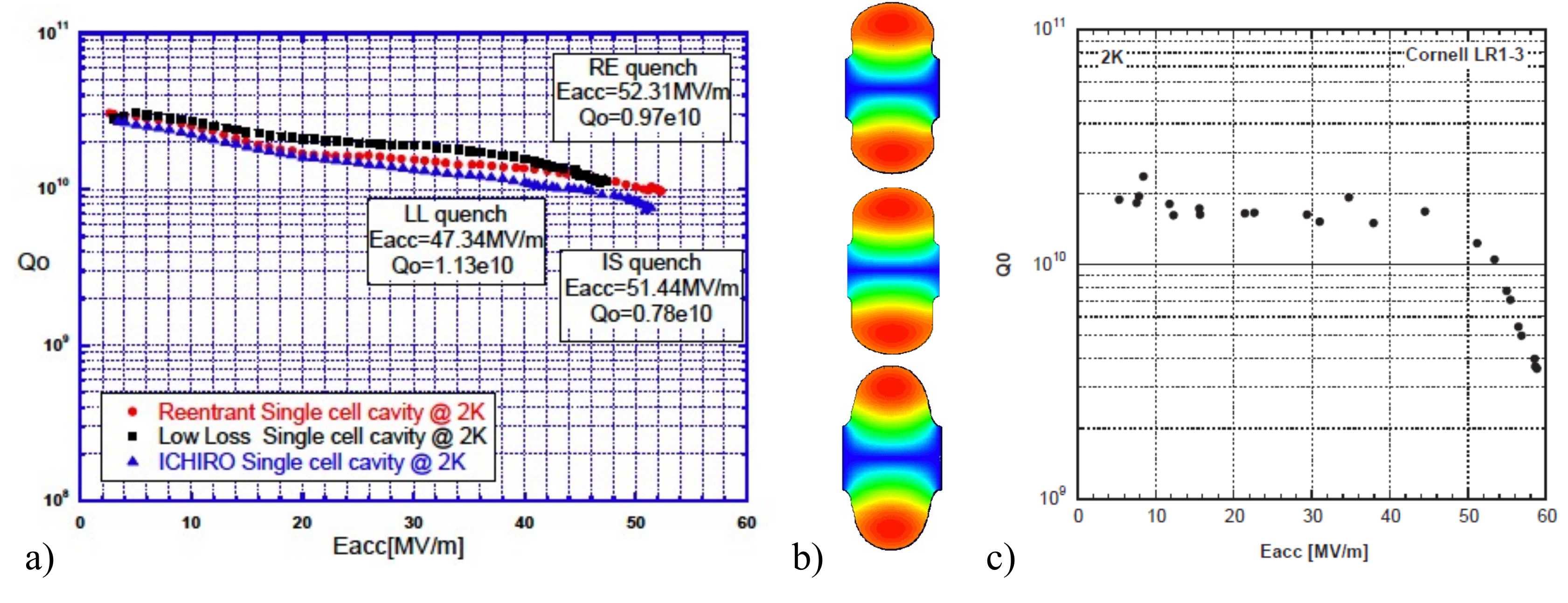}
    \caption{(a) Gradients greater than 50~MV/m demonstrated in single-cell cavities of various improved shapes \cite{Cavity_shapes}. (b) Comparison of Re-entrant (RE, top), Low Loss/ICHIRO (middle) and TESLA (bottom) cavity cell shapes, showing the magnetic field intensity with red highest and blue lowest. (c) Record gradient near 59 MV/m demonstrated with the re-entrant shape (60~mm aperture) \cite{Re-entrant_record}.}
    \label{fig:Cavity_shapes}
\end{figure}

Since peak surface magnetic field $H_{pk}$ presents a hard ultimate limit to the performance of Nb cavities via the critical superheating field, several new cavity geometries were developed to improve upon the TESLA cavity geometry. Re-entrant \cite{Re-entrant_cavity, Optimal_cells}, Low-Loss \cite{Low-Loss_cavity} and ICHIRO \cite{ICHIRO_cavity} shapes (Figure~\ref{fig:Cavity_shapes}(b)) have been introduced to lower the $H_{pk}/E_{acc}$ ratio by 10--20\% via rounding the equator to expand the surface area of the high magnetic field region, and by allowing $E_{pk}/E_{acc}$ to rise by about 20\%. The $G\cdot R/Q$ value\footnote{This figure of merit depends only on the cavity shape and not its frequency/size \cite{Padamsee_book_II}. It allows comparing efficiency of different geometries by separating the geometry-only parameters from the surface resistance $R_s$ of cavity material. $G$ is the geometry factor and $R/Q$ is the specific shunt impedance. For a given cavity voltage $V_c$, the power dissipated in the cavity walls is given by $P_c = V_{c}^{2} \cdot R_s / (G\cdot R/Q) = V_{c}^{2} / R_{sh}$. $R_{sh}$ is the cavity shunt impedance.} for the advanced shapes is about 30--40\% higher than the TESLA shape, which reduces cryogenic losses. The 20\% increase in $E_{pk}$ makes cavities with the new shapes more susceptible to field emission, but we can expect progress in field emission reduction with cleaner surface preparation developments, as well as plasma processing over the coming years.

Many single cell cavities with the advanced shapes were built, prepared with the standard ILC treatment recipe, and tested to demonstrate gradients of 50--54~MV/m with $Q_0$ values above $10^{10}$, as shown in Figure~\ref{fig:Cavity_shapes}(a) for cavities tested at KEK \cite{Cavity_shapes}. A record field of 54~MV/m at $Q$ about $10^{10}$ was set by a single cell Re-entrant cavity with 60~mm aperture, and 59~MV/m at $Q$ about $3\times10^9$ (see Figure~\ref{fig:Cavity_shapes}(c) \cite{Re-entrant_record}) for the same cavity. However, the best multi-cell cavities of the new shapes have only reached 42~MV/m \cite{Multi-cell_new_shapes}, mostly due to the dominance of field emission. A relative newcomer to the advanced shape effort is the Low Surface Field (LSF) shape \cite{Li_LSFcavity} which obtains $H_{pk}/E_{acc}$ of 37.1~Oe/(MV/m), as compared to 42.6~Oe/(MV/m) for TESLA cavity, without raising $E_{pk}/E_{acc}$ (= 1.98). It is very similar to the Low-Loss shape. The LSF shape may be the immediate answer to the field emission challenges of the advanced shapes. A 9-cell cavity at JLAB has demonstrated accelerating gradient of >45~MV/m in four cells including 51~MV/m in two cells, as determined by powering the cavity in several modes of the fundamental passband \cite{Geng_IPAC2021}.

As we have seen earlier, the newly developed, two-step bake procedure has demonstrated a gradient of 49~MV/m in TESLA shape 1-cell cavities. Combining the two-step bake with one of the advanced shape cavities has the potential of improving the gradients toward 60~MV/m. For example, the Low-Loss shape has the potential for 18\% improvement from 49 to 58~MV/m. But no laboratory has attempted such combined efforts as yet.

Note that the SRF acceleration structure is a self-consistent system, which contains in addition to the acceleration cavity different components: frequency tuners, high-power coupler, helium vessel, and so on. Increase in acceleration gradient imposes new requirements, and to accommodate these requirements it is necessary to either upgrade the components or develop new ones. Below we give two examples.

The first example provides justification for developing new cavity resonance control systems. At a fixed beam current $I_{beam}$ the cavity half-bandwidth (HBW) is 
\begin{equation}
    \delta f_{HBW} = f \frac{R/Q \cdot I_{beam}}{2E_{acc}L} ,
\end{equation}
where $f$ is the cavity resonance frequency and $L$ the cavity length. On the other hand, the cavity Lorenz Force Detuning (LFD) is
\begin{equation}
    \Delta f_{LFD} = k E_{acc}^2 ,
\end{equation}
where $k$ is the LFD coefficient. For TESLA cavity it is $\sim1$~Hz/(MV/m)$^2$. This means that 
\begin{equation}
    \frac{\Delta f_{LFD}}{\delta f_{HBW}} \sim E_{acc}^3 .
\end{equation}
For high gradients $E_{acc} > 50$~MV/m this ratio may be higher than 30--40, which creates problems for stable cavity operation in a pulsed mode.
To mitigate LFD in a high-gradient SRF linear collider, new types of passive and active means should be developed including new resonance control algorithms.

The second example outlines the need for high-power RF components. The higher accelerating gradient is, the higher is the demand for input RF power:
\begin{equation}
    P = I_{beam}E_{acc}L \sim E_{acc} ,
\end{equation}
and therefore, new Fundamental Power Couplers (FPC) must be developed to accommodate higher pulsed and average power requirements. In addition, to satisfy the higher cavity input power requirement, we may need to develop new RF sources and RF power distribution systems.

\subsubsection{Nb-based traveling wave SRF}
\label{sec:TW_SRF}

Travelling wave (TW) structures offer several main advantages compared to standing wave structures: substantially lower peak magnetic ($H_{pk}/E_{acc}$), lower peak electric field ($E_{pk}/E_{acc}$) ratios, together with substantially higher $R/Q$ (for lower cryogenic losses). In addition, TW structure provides high stability of the field distribution along the structure with respect to geometrical perturbations. This allows for a much longer accelerating structures than TESLA cavities, limited by manufacturing technology.

\begin{figure}
    \centering
    \includegraphics[width=0.8\textwidth]{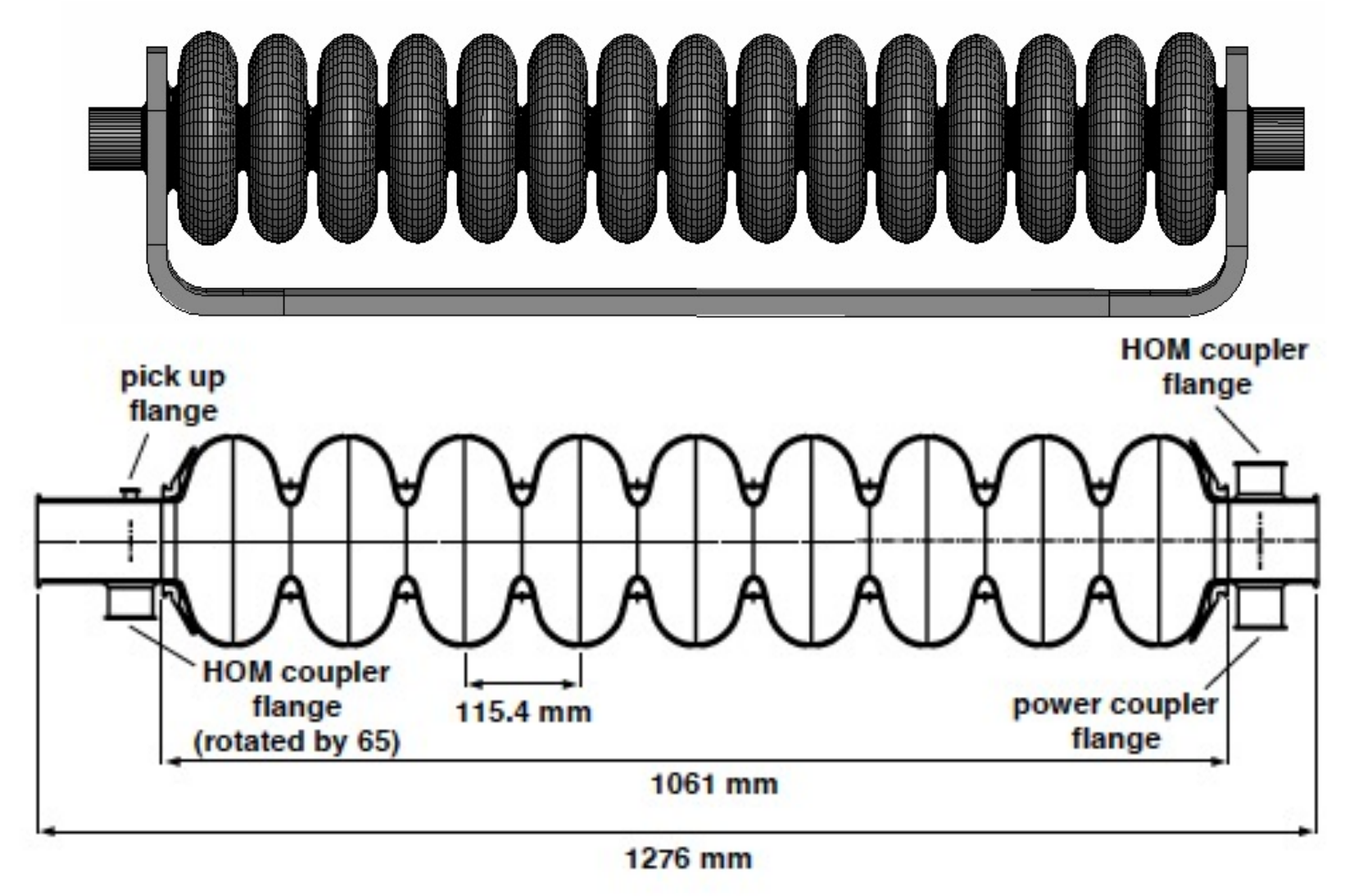}
    \caption{The TW structure with a 105$^\circ$ phase advance per cell compared to the one-meter standing-wave TESLA structure \cite{TESLAcavity}.}
    \label{fig:TW_vs_SW}
\end{figure}

The emphasis for future design is to lower $H_{pk}/E_{acc}$, as much as possible, since $H_{pk}$ presents a hard ultimate limit to the performance of Nb cavities via the critical superheating field. But, as Figure~\ref{fig:TW_vs_SW} shows, the TW structure requires almost twice the number of cells per meter as for the SW structure in order to provide the proper phase advance (about 105 degrees), as well as a feedback waveguide for redirecting power from the end of the structure back to the front of the accelerating structure, which avoids high peak surface fields in the accelerating cells. The feedback requires careful tuning to compensate reflections along the TW ring to obtain a pure traveling wave regime at the desired frequency.

\begin{figure}[h]
    \centering
    \includegraphics[width=0.25\textwidth]{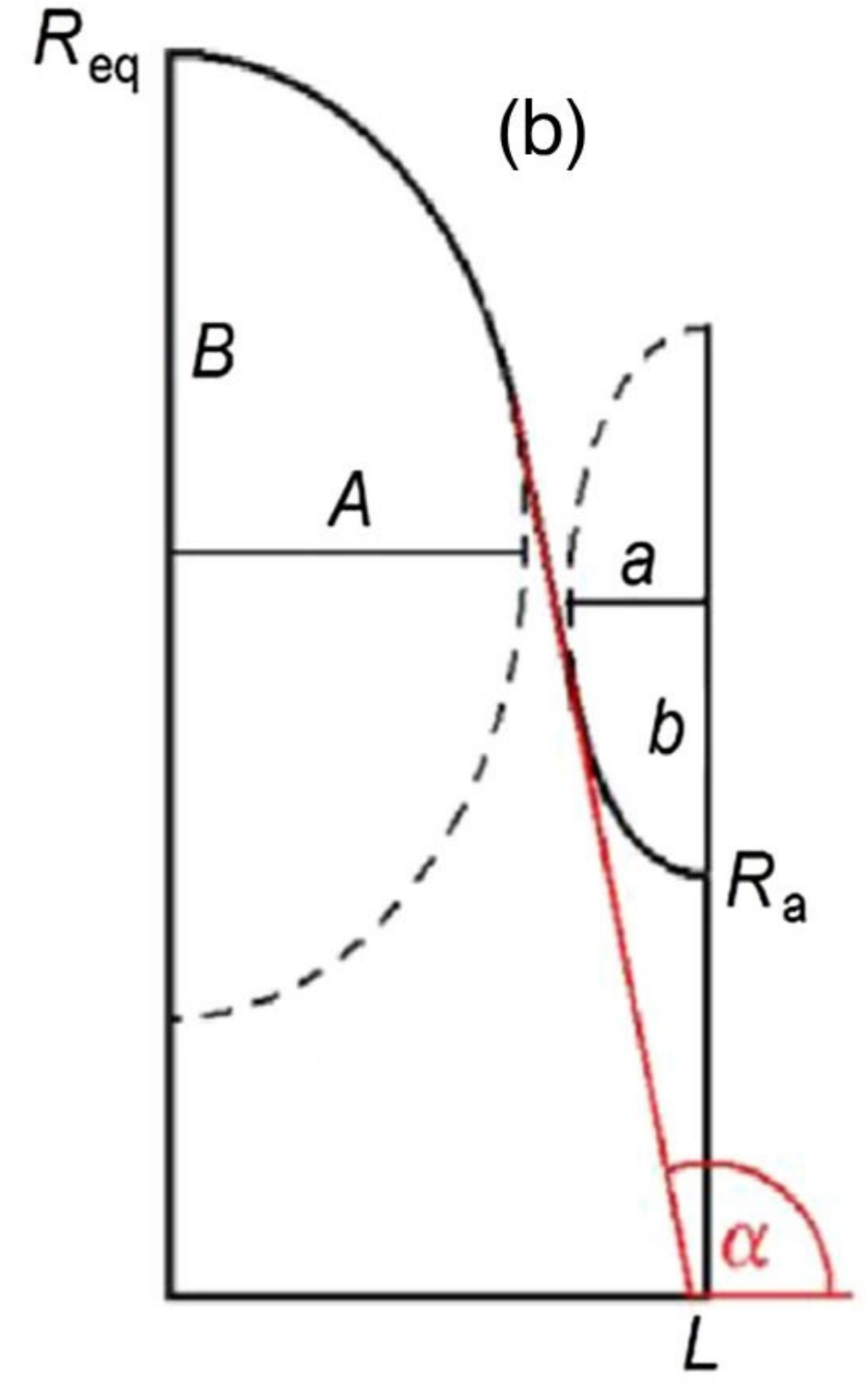}
    \caption{Geometry of the TW half cell.}
    \label{fig:TW_geometry}
\end{figure}

Table~\ref{tab:TW_parameters} \cite{TW_optimization} shows one set of parameters for optimized cell shape, phase advance, and 50~mm aperture that yield $H_{pk}/E_{acc}$ = 28.8~Oe/(MV/m) with $E_{pk}/E_{acc}$ = 1.73. The geometrical parameters for the cell shape are defined in Figure~\ref{fig:TW_geometry}.  Since $H_{pk}/E_{acc}$ is 42.6~Oe/(MV/m) for the TESLA structure, the TW structure has reduced the critical parameter $H_{pk}/E_{acc}$ by a factor of 1.48! If results for the best single cell TESLA shape cavities prepared today ($E_{acc}$ = 49~MV/m, $H_{pk}$ = 2090~Oe) can be reached in such a TW structure, we can optimistically expect a gradient $E_{acc}$ > 70~MV/m. 

\begin{table}
\begin{center}
\begin{tabular}{l c}
     
\hline
\hline
Optimization & 120/200 \\
\hline
Phase advance (deg.)  & 90 \\
$A$ (mm) & 23.83 \\
$B$ (mm) & 36.40 \\ 
$a$ (mm) & 4.51 \\
$b$ (mm) & 7.52 \\
$E_{pk}/E_{acc}$ & 1.727 \\
$B_{pk}/E_{acc}$ (mT/(MV/m)) & 2.878 \\
$R_{sh}/Q$ (Ohm/m) & 2,127 \\
$\alpha$ (deg.) & 90.91 \\
$R_{eq}$ (mm) & 98.95 \\
$v_{gr}/c$ & 0.01831 \\
$E_{acc}$ (MV/m) & 69.5 \\
$E_{acc}\cdot2L$ MV & 4.00 \\
\hline 
\end{tabular}
\end{center}
\caption{Parameters of optimized cells with limiting surface fields $E_{pk}$ = 120~MV/m and $B_{pk}$ = 200~mT; $L − A$ = 5~mm, aperture radius $R_a$ = 25~mm (from \cite{TW_optimization}, Table~III, column 3.) Geometrical parameters are shown in Figure~\ref{fig:TW_geometry}.}
 \label{tab:TW_parameters}
\end{table}

The 100\% $R/Q$ increase (even with somewhat lower $G$ of 186 Ohm vs. 270 Ohm for TESLA) lowers the dynamic heat load and cryogenic power needed for high gradients. The high group velocity in the TW mode also increases the cell-to-cell coupling from 1.8\% for the TESLA structure to 2.3\%. Thus, TW structures have less sensitivity to cavity detuning errors, making tuning easier, despite the larger number of cells. Studies \cite{TW_optimization} show that the cell shape can be fine-tuned to avoid multipacting, without increasing $H_{pk}$ more than 1\%. Higher Order Mode (HOM) damping for TW structures is under study. Preliminary results show that the first 10 monopole modes up to 7~GHz show no trapping.

Many significant challenges must still be addressed along the TW development path. High circulating power in the feedback waveguide must be demonstrated. Cavity fabrication and surface processing procedures and fixtures must deal with (roughly) double the number of cells per structure.

\begin{figure}
    \centering
    \includegraphics[width=0.8\textwidth]{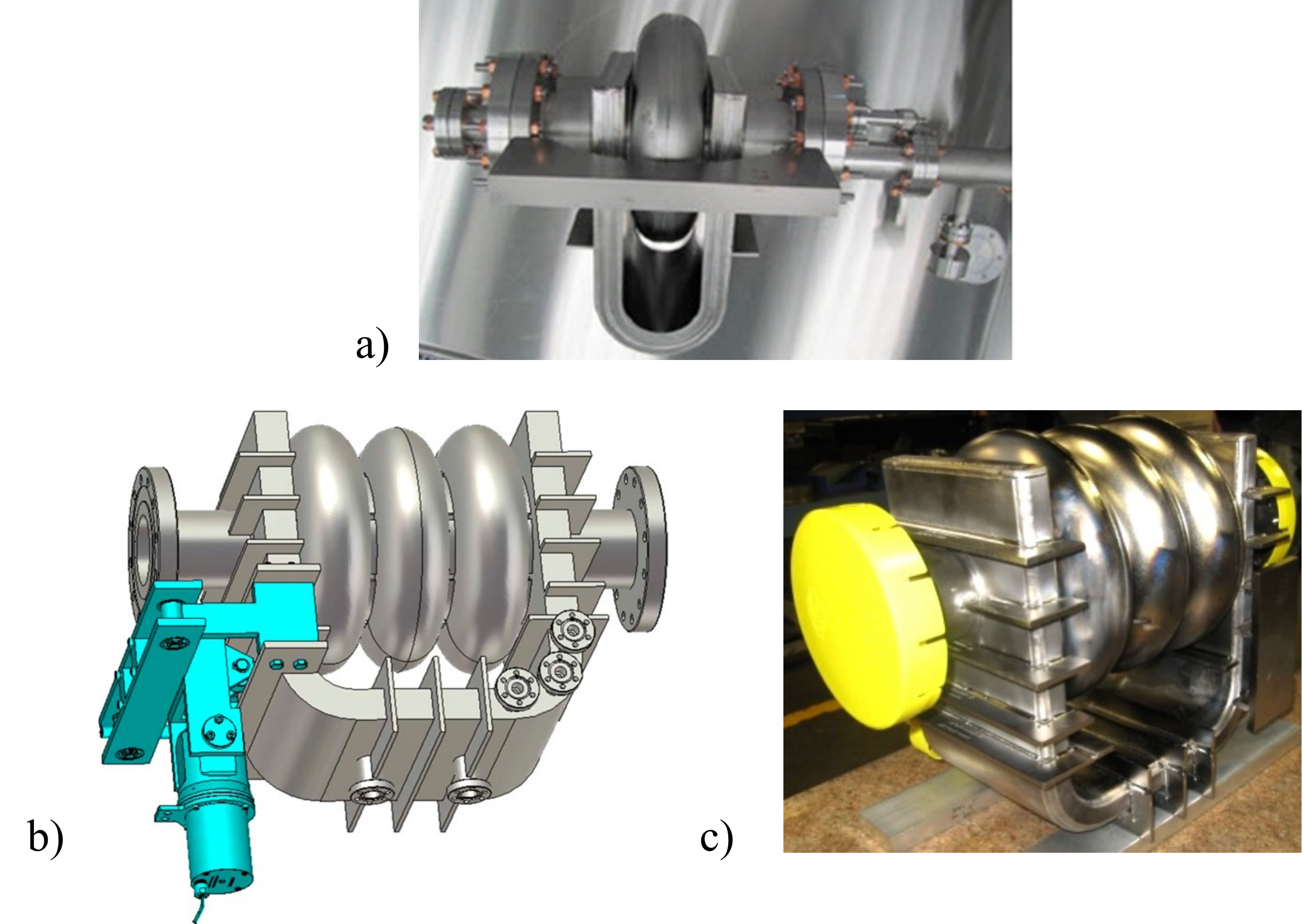}
    \caption{(a) Single-cell TW niobium structure with feedback waveguide, treated by BCP and tested to reach 26~MV/m; (b) 3D model of the 3-cell TW structure; (c) Fabricated 3-cell cavity.}
    \label{fig:TW_proof-of-principle}
\end{figure}

First structure fabrication and testing efforts have started for TW cavity development \cite{Kostin_1-cell_test}. With the relatively easier BCP treatment only, the first single cell TW cavity (Figure~\ref{fig:TW_proof-of-principle}(a)) with recirculating waveguide achieved 26~MV/m accelerating gradient, limited by the high field $Q$-slope, as expected for BCP. This result is very encouraging for the first attempt. A 3-cell Nb TW structure with recirculating waveguide, shown in Figure~\ref{fig:TW_proof-of-principle}(b, c), was designed and fabricated but has not yet been tested \cite{Kostin_3-cell_progress}.

\subsubsection{Advances in Nb$_3$Sn and other materials}
Nb$_3$Sn is a promising new material for SRF cavities. In R\&D efforts, it has already shown high $Q\sim10^{10}$ even at relatively high temperatures $\sim$4~K. Single cell cavities have reached 24~MV/m and 9-cell TESLA cavities have reached 15 MV/m \cite{Posen_2021}. However, the full potential of Nb$_3$Sn is significantly higher. The highest fields reached in typical niobium cavities, $\sim$50~MV/m \cite{2-step_baking}, is consistent with being limited by the superheating field of niobium \cite{Hsh1,Hsh2,Hsh3}. Based on its superconducting properties, the corresponding limit for Nb$_3$Sn would be approximately twice as high, $\sim~$100~MV/m. 

\begin{figure}
    \centering
    \includegraphics[width=0.8\textwidth]{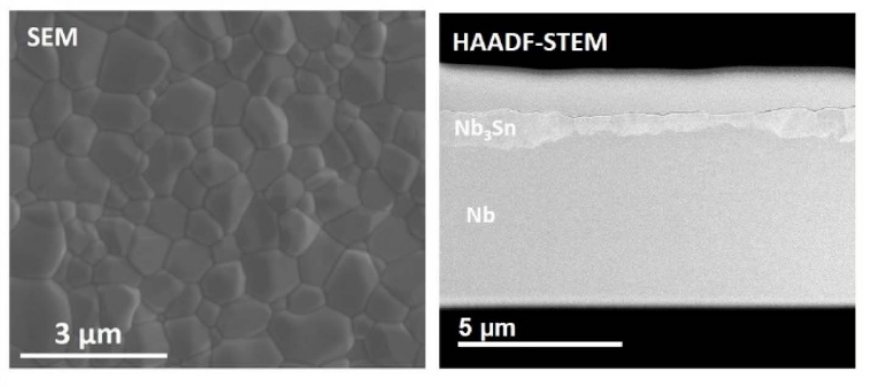}
    \caption{Top view and cross section of a Nb$_3$Sn SRF film. From \cite{Posen_2021}.}
    \label{fig:Nb3SnMicroscopy}
\end{figure}

Experiments suggest that Nb$_3$Sn cavities are limited by defects in the RF surface \cite{PosenPRL,PorterThesis}. Nb$_3$Sn is expected to be more sensitive to surface defects than niobium because of its relatively short coherence length $\sim$3--4 nm, approximately an order of magnitude smaller than niobium, depending on the niobium surface treatment. On the other hand, Nb$_3$Sn R\&D on surface treatments after coating is still relatively primitive. Coated surfaces are relatively rough on relevant length scales (see Figure~\ref{fig:Nb3SnMicroscopy}), and attempts to smoothen surfaces using techniques developed for niobium have so far resulted in other issues such as residues and performance degradation \cite{PosenThesis,JLabNb3SnEP}. 

\begin{figure}
    \centering
    \includegraphics[width=0.7\textwidth]{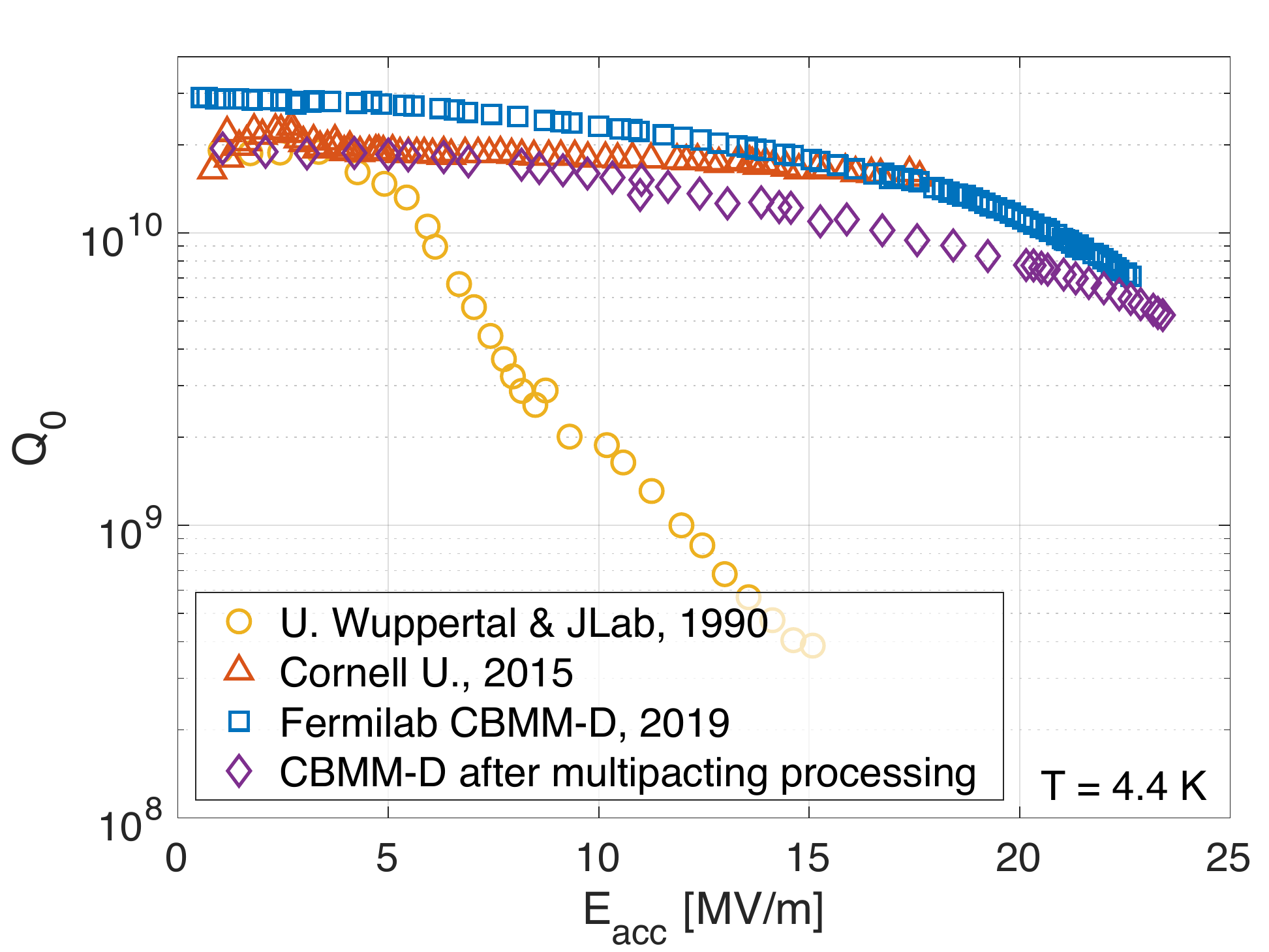}
    \caption{Progress in performance of 1.3–1.5 GHz single cell cavities at 4.4 K (from \cite{Posen_2021}).}
    \label{fig:Nb3Snprogress}
\end{figure}

However, compared to niobium, relatively little effort has so far been invested in surface processing of Nb$_3$Sn. New R\&D is underway to smoothen Nb$_3$Sn surfaces and also make the coatings thinner, which could help to thermally stabilize any defects that are present by reducing thermal impedance. There are promising directions, including electropolishing, oxypolishing, and mechanical polishing, as well as new deposition methods to try to create inherently smoother films. The Nb$_3$Sn R\&D efforts are funding limited, but even with relatively small efforts, there have been significant advances in Nb$_3$Sn performance over the last years, as shown in Figure~\ref{fig:Nb3Snprogress}. Further support could help to explore the full potential of this material. Reaching even 60--70\% of the full potential of Nb$_3$Sn would be extremely helpful for enabling the HELEN collider.

%% file: HELENcollider.tex
\section{HELEN collider}
\label{sec:HELENcollider}

For more than four decades, efforts have been devoted to developing high-gradient RF technology linear $e^+e^-$ colliders in order to overcome the synchrotron radiation limitations of circular  $e^+e^-$ machines. In linear colliders, where each bunch collides only once, the primary challenge confronting high luminosity is the beam power requirement \cite{shiltsev2021modern}: 

\begin{equation}
\label{eq:lumi4}
 {\cal L} = \frac{1}{8\pi \alpha r_0}\,\frac{P_{\rm wall}}{\sqrt{s}} 
\, \frac{\eta}{\sigma_y^{\ast}} \; N_{\gamma}\, H_{\rm D}\, . 
\end{equation}
Here, $P_{\rm wall}$ is the total wall-plug power of the collider, to be converted into beam power $P_{\rm b}=2 f_0 N E_{\rm b}$ with efficiency $\eta$,  $N_{\gamma}\approx 2 \alpha r_0 N/\sigma_{x}^{\ast}$ is the number of beamstrahlung photons emitted per $e^{\pm}$ ($\alpha$ denotes the fine-structure constant), and the last factor $H_{\rm D}$, typically between 1 and 2, represents the enhancement of luminosity due to the {\it pinch  effect},  the additional focusing occurring during the collision of oppositely charged bunches. To maximize the luminosity with fixed $P_{\rm wall}$, one has to push the bunch population $N$ and cope with beamstrahlung (radiation of photons due to the electromagnetic field of opposing bunch). 

Beamstrahlung is very much an issue for all linear colliders as it may significantly widen the luminosity center of mass energy (c.m.e.) spectrum, especially at higher c.m.e. above $\sim$500~GeV. The effect is mitigated by making the colliding beams as flat as possible at the interaction point ($\sigma_{x}^\ast \gg \sigma_{y}^\ast$).  The photon energy spectrum of the beamstrahlung is characterized by the parameter $\Upsilon= (2/3) \hbar \omega_{\rm c}/E_{\rm b}$ \cite{accel:Yokoya1991qz}, with $\hbar \omega_{\rm c}$ denoting the critical photon energy and $E_{\rm b}$ the beam energy.  The spectrum strongly deviates from the classical synchrotron radiation spectrum for $\Upsilon$ approaching or exceeding 1. Keeping a significant fraction of the luminosity close to the nominal energy represents a design goal, which is met  if $N_{\gamma}$ does not exceed a value of about 1.  A consequence is the use of flat beams, where $N_\gamma$ 
is managed by the beam width, and luminosity adjusted by the beam height, 
thus the explicit appearance of the vertical beam size $\sigma_y^{\ast}$. 

The strength of the EM beam-beam interaction at the IP of linear colliders is determined by the  {\it disruption parameter} $D_{y}$ -- the ratio  of the rms bunch length $\sigma_{z}$ to the beam-beam focal length -- related to the beam-beam parameter $\xi_{y}$ via $D_{y}= 4  \pi \sigma_{z} \xi_{y}/\beta^*_{y}$. Significant disruption leads to effectively smaller beam size due to {\it traveling focus} effect, and a resulting luminosity enhancement. It also makes the collision more  sensitive to small offsets, resulting in a {\it kink instability}. Additional effects arising include $e^+e^-$ pair creation, and  depolarization by various mechanisms. 

In Table~\ref{tab:LCtable} we compare parameters of HELEN with other $e^{+}e^{-}$ linear colliders. Note that this tentative baseline set of parameters for HELEN collider is for the TW SRF linac option. As mentioned in section~\ref{sec:TW_SRF}, high stability of field distribution along the TW structure allows for longer structures. For baseline HELEN parameters we assumed the active cavity length of 2.37~m, about 2 times longer than TESLA cavity. Other collider options will be discussed in section~\ref{sec:HELENparameters}. As it is obvious from the table, HELEN is in many aspects a high-gradient modification of the the International Linear Collider (ILC). The ILC has a baseline center-of-mass energy of 500~GeV or 250~GeV for a Higgs factory option \cite{ILC_Snowmass2021}. The ILC is based on the 1.3 GHz TESLA superconducting accelerating structures with 31.5 MV/m average gradient, and aims at 7.7 nm vertical beam size at the IP. Several scenarios of luminosity and energy upgrades of the ILC are under consideration \cite{ILC_Snowmass2021, padamsee2021ilc}. The ILC luminosity upgrade scenarios are directly applicable to HELEN. Possible energy upgrade options we consider in section~\ref{sec:HELENparameters}.

CERN's Compact Linear Collider (CLIC) design \cite{CLICreference}, under development since the mid-1980s,  also includes possible upgrades, from an initial 380~GeV c.m.e. to ultimately 3~TeV, which would enable searches for new particles of significantly higher masses. CLIC is based on a novel two-beam  acceleration scheme in which normal conducting (NC) copper high-gradient 12~GHz accelerating structures are powered by a high-current 1.9~GeV drive beam to enable accelerating gradients up to 100~MV/m (though optimal gradient for the first CLIC stage at $\sqrt{s}$ = 380~GeV is 70~MV/m, and for this stage an alternative RF power drive option with 12 GHz klystrons powering is also being considered).

Recent "Cool Copper Collider" (C$^3$) proposal \cite{nanni2021c} envisions klystron-driven C-band NC cavities operating at liquid nitrogen temperatures to achieve higher shunt impedance, smaller breakdown rate, and 70--120~MV/m accelerating gradients. 

All linear colliders have a lot in common, e.g., their main subsystems include damping rings, and sophisticated beam delivery (final focus) systems. To reach their design luminosities, linear colliders require very high rates of positron production, and very tight control of imperfections, such as  $O$(10~$\mu$m) accuracy of pre-alignment of the main linac and beam delivery system components, suppression of fast vibrations of the quadrupoles due to ground motion to $O$(1~nm) level at frequencies above 1~Hz, advanced beam-based trajectory tuning, and  mitigation of wakefield effects.

\begin{table}
\begin{center}
\begin{adjustbox}{angle=90}

\begin{tabular}{| l || c || c | c | c |}
     
\hline
\hline
 Parameter & HELEN & C$^3$ & ILC & CLIC \\ 
\hline
CM energy 2$\times E_{\rm b}$ (GeV)&
	{250}&
	{250, 550}&
       	{250, 500}&
	{380, 3000} \\
Length (km) &
	{7.5} &
	{8, 8} &
	{20.5, 31} &
        {11.4, 50} \\
Interaction points  &
	{1} &
        {1} &
       	{1} &
	{1} \\ 
Integrated luminosity (ab$^{-1}$/yr) & 
	{0.2}&
	{0.2, 0.4}&
        {0.2, 0.3}&
        {0.1, 0.6} \\ 
Peak lumi. $\cal L$ (10$^{34}$cm$^{-2}$s$^{-1}$) &
	{1.35} &
	{1.3, 2.4} &
	{1.35, 1.8} &
	{1.5, 6 } \\
CM energy spread $\sim 0.4\delta_{\rm BS}$ (rms, \%) &
	1 &
	1.6, 7.6 &
	1, 1.7 &
	1.7, 5 \\ 
Polarization (\%)&
	\begin{tabular}[c] {@{}c@{}} 80/30  ($e^-/e^+$)  \end{tabular} &
        {tbd}&
	\begin{tabular}[c] {@{}c@{}} 80/30  ($e^-/e^+$)  \end{tabular} &
	\begin{tabular}[c] {@{}c@{}} 80/0  ($e^{-}/e^{+}$) \end{tabular} \\ 
\hline
Rep.rate $f_{\rm rep}$ (Hz) &
	{5} &
	{120} &
	{5} &
	{50} \\ 
Bunch spacing (ns) &
	{554} &
	{5.26, 3.5} &
	{554} &
	{0.5} \\ 
Particles per bunch $N$ ($10^{10}$) &
	{2}&
	{0.63}&
	{2}&
        {0.52, 0.37 } \\
Bunches per pulse $n_{\rm b}$ &
	{1312}&
	{133, 75}&
        {1312}&
        {352, 312 } \\
Pulse duration ($\mu$s)  &
	{727}&
	{0.7, 0.26}&
	{727}&
	{0.176, 0.156} \\
Pulsed beam current $I_{\rm b}$ (mA)  &
	{5.8}&
	{190, 286}&
	{5.8}&
	{1670, 1190} \\
\hline
Bunch length $\sigma_z$ (rms, mm)&
	{0.3}&
	{0.1}&
	{0.3}&
	{0.07, 0.044} \\ 
IP beam size $\sigma^*$ (rms, $\mu$m)&
	\begin{tabular}[c] {@{}c@{}} H: 0.52 \\ V: 0.0077 \end{tabular}&
	\begin{tabular}[c] {@{}c@{}} H: 0.23, 0.16    \\ V: 0.004, 0.0026          \end{tabular}&
	\begin{tabular}[c] {@{}c@{}} H: 0.52, 0.47  \\ V: 0.0077, 0.0059       \end{tabular}&
	\begin{tabular}[c] {@{}c@{}} H: 0.15, 0.04  \\ V: 0.003, 0.001       \end{tabular}\\ 
Emittance, $\varepsilon_{\rm n}$ (rms, $\mu$m)&
	\begin{tabular}[c] {@{}c@{}} H: 5  \\ V: 0.035 \end{tabular} &
	\begin{tabular}[c] {@{}c@{}} H: 0.9  \\ V: 0.02 \end{tabular} &
	\begin{tabular}[c] {@{}c@{}} H: 5, 10  \\ V: 0.035, 0.035 \end{tabular} &
	\begin{tabular}[c] {@{}c@{}} H: 0.95, 0.66  \\ V: 0.03, 0.02 \end{tabular} \\ 
$\beta^*$ at interaction point (mm)&
	\begin{tabular}[c] {@{}c@{}} H: 13        \\ V: 0.41
	\end{tabular} &
	\begin{tabular}[c] {@{}c@{}} H: 12      \\ V: 0.12 
	\end{tabular} &
	\begin{tabular}[c] {@{}c@{}} H: 13, 11        \\ V: 0.41, 0.48 \end{tabular} &
	\begin{tabular}[c] {@{}c@{}} H: 8, 6.9      \\ V: 0.1, 0.068 \end{tabular} \\ 
Full crossing angle $\theta_{\rm c}$ (mrad)&
	{14} &
	{14 } &
	{14}&
	{20} \\
Crossing scheme&
	{crab crossing} &
	{crab crossing} &
	{crab crossing }&
	{crab crossing } \\
Disruption parameter $D_{y}$ &
        {35}&
        {12}&
         {35, 25}&
         {13, 8} \\
\hline
RF frequency $f_{\rm RF}$ (MHz)&
	{{{1300}}}&
	{{{5712}}}&
	{{{1300}}}&
	{{{11994}}} \\
Accelerating gradient $E_{acc}$ (MV/m)  &
	{70}&
	{70, 120}&
        {31.5}&
        {72, 100 } \\ 
Effective gradient $E_{eff}$ (MV/m)  &
	{55.6}&
	{63, 108}&
        {21}&
        {57, 79 } \\ 
Total beam power (MW) &
	{5.3}&
	{4, 4.9}&
	{5.3, 10.5}&
	{5.6, 28 } \\
Site power (MW) &
	{110}&
	{$\sim$150, $\sim$175}&
	{111, 173} &
	{168, 590} \\
Key technology &
	{{{TW SRF}}}&
	{{{cold NC RF}}}&
	{{{SW SRF}}}&
	{{{two-beam accel.}}} \\
\hline 
\end{tabular}

\end{adjustbox}

\end{center}
\caption{Tentative parameters of HELEN and other $e^+e^-$ linear collider Higgs factory proposals. Parameters associated with different beam energy scenarios are comma-separated; H and V 
indicate horizontal and vertical directions; a fill factor of 80.4\% is assumed to calculate the real-estate (effective) gradient $E_{eff}$ of HELEN.}
\label{tab:LCtable}
\end{table}

\subsection{Emittance preservation and luminosity degradation control}
Main sources leading to emittance growth and/or luminosity degradation in linear colliders are: a)~wakefield effects, primarily in accelerating RF structures but also from other apertures such as collimators; b) chromatic (i.e., dispersive) effects, arising from magnet misalignment and beam trajectory errors; and c) beam-beam separation at the IP due to jitter of the final focus quadrupoles. 

HELEN, as the ILC, has quite relaxed (compared to other linear collider proposals) alignment and jitter tolerances due to two factors: a) large aperture L-band superconducting RF cavities having the relatively low cavity wakefields and mechanically alignment accuracy of the accelerating cavities $O$(300~$\mu$m) is good enough for suppressing single-bunch wakefield effect to acceptable levels; and b) long pulse train and significant bunch spacing that allows to measure position of the first few bunches and apply necessary corrections to keep the rest of the train on right trajectory.  Beam Position Monitors (BPMs) accuracy $O$(1~$\mu$m) is typically sufficient for the 
beam-based feedback system that includes: a) a slow feedback correcting the beam orbit to compensate for low frequency ground motion; b) an inter-pulse feedback acting in a few locations to correct accumulated errors that occur in between the action of the slow system and to provide the possibility of straightening the beam; and c) a fast intra-train
feedback system acting at the IP to keep the beams in collision, correcting for the high frequency ground motion that moves the final quadrupole doublet. 
Fermilab has significant expertise in the area of collider element stabilization and emittance control, and several important experimental studies indicate that Fermilab's site is sufficiently quiet for an L-band linear collider \cite{baklakov1998ground, shiltsev2010review}.

\subsection{Parameters, layout, siting, and upgrades}
\label{sec:HELENparameters}

\begin{figure}
    \centering
    \includegraphics[width=\textwidth]{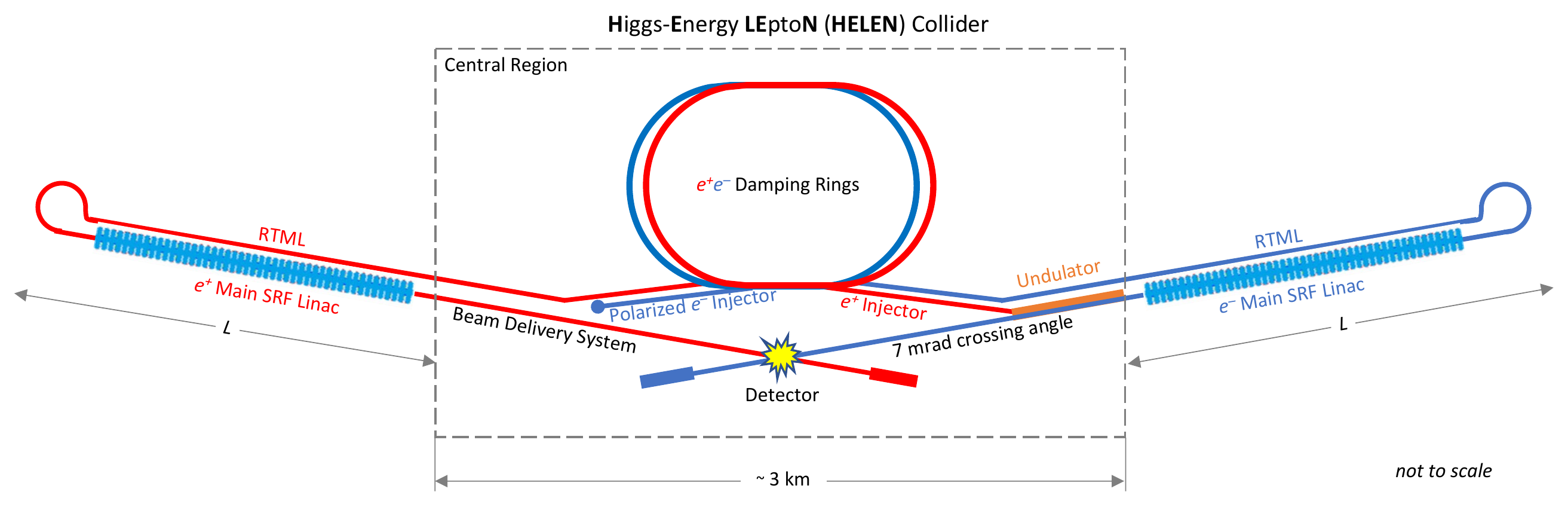}
    \caption{Conceptual layout of the HELEN collider.}
    \label{fig:HELEN_layout}
\end{figure}

The baseline SRF technology option for HELEN is the traveling wave accelerating structure described in \ref{sec:TW_SRF} with tentative parameters given in Table~\ref{tab:LCtable}. This option provides an optimal combination of the high accelerating gradient of 70~MV/m with an expected demonstration of a fully developed cryomodule within $\sim5$~years, providing that sufficient funding is available. In this section we would like to compare HELEN parameters for three different options listed in the order of their technical difficulty. In the following we assume that the length occupied by the beam delivery system (BDS) is 3~km. The layout of the collider (Figure~\ref{fig:HELEN_layout}) is similar to that of the ILC. The only difference between the three option is the length $L$ of main SRF linacs.

\textbf{Option 1: Advanced geometry SW structure operating at 55 MV/m.} 
As we mentioned in section~\ref{sec:SW_SRF}, combing and advanced cavity shape and new treatment recipes should allow reaching accelerating gradients of $\sim 60$~MV/m. This version is essentially the ILC with different SRF cavities operating at higher gradient. Assuming the LSF accelerating structure operating at 55~MV/m and the fill factor of 71\%, the HELEN collider would be 9.4~km long. 

\textbf{Option 2 (baseline): TW structure operating at 70 MV/m.}
The traveling wave option assumes an accelerating gradient of 70~MV/m. With accelerating structures about 2 times longer than TESLA cavities, the fill factor increases to 80.4\% and the collider will be 7.5 km long. 

\textbf{Option 3: Nb$_3$Sn structure operating at 90 MV/m.} For this option, we assume the LSF-shape cavities operating at 90 MV/m at $\sim 4$~K. This shortens the collider length to 6.9 km.
\vspace{3mm}

\noindent
Some parameters of the three options are compared in Table~\ref{tab:HELENtable}. The TW option was selected as the baseline for the proposed HELEN collider for the following reasons:
\begin{itemize}
    \item It is the most efficient in terms of AC power consumption and is on par with the ILC site power demand (see Table~\ref{tab:LCtable}). 
    \item It offers the best cost saving/ Our preliminary cost estimate indicate that the cost savings (relative to the ILC main linac cost) are 13\% for Option 1, 26\% for Option 2, and 18\% for Option 3.
    \item The traveling wave technology can be demonstrated on a relatively short time scale. With an aggressive R\&D program and innovations in cavity surface treatments and processing, the required accelerating gradient can possibly be reached within the next 2--3 years. After that, another 2--3 years would be needed to build and test a demonstration cryomodule, possibly with beam at the Fermilab's FAST facility.
\end{itemize}

\begin{table}
\begin{center}

\begin{tabular}{| l || c | c | c | }
     
\hline
\hline

Parameter & Advanced SW & Traveling wave & Nb$_3$Sn \\ 
\hline
Accelerating gradient (MV/m)&
	55&
	70&
	90 \\
Fill factor &
    0.711 &
    0.804 &
    0.711 \\
Real estate (effective) gradient (MV/m) &
    39.1 &
    55.6 &
    64.0 \\
Cavity $Q$ ($10^{10}$)  &
	1.0 (2 K) &
        0.69 (2 K) &
       	1.0 (4.5 K) \\ 
Active cavity length (m) &
    1.038 &
    2.37 &
    1.038 \\
Cavity $R/Q$ (Ohm) &
    1158 &
    4890 &
    1158 \\
Geometry factor $G$ (Ohm) &
    279 &
    186 &
    279 \\
$B_{pk}/E_{acc}$ mT/(MV/m) &
    3.71 &
    2.89 &
    3.71 \\
$E_{pk}/E_{acc}$ &
    1.98 &
    1.73 &
    1.98 \\
Number of cavities &
    4380 &
    1527 &
    2677 \\
Number of cryomodules &
    505 &
    382 &
    309 \\
Collider length (km) &
	9.4 &
	7.5 &
	6.9 \\
AC power for main linacs (MW) &
    49 &
    39 &
    58 \\
Total collider AC power (MW) &
    121 &
    110 &
    129 \\
\hline 
\end{tabular}

\end{center}
\caption{Comparison of some HELEN collider parameters for three option.}
\label{tab:HELENtable}
\end{table}

\begin{figure}
    \centering
    \includegraphics[width=0.7\textwidth]{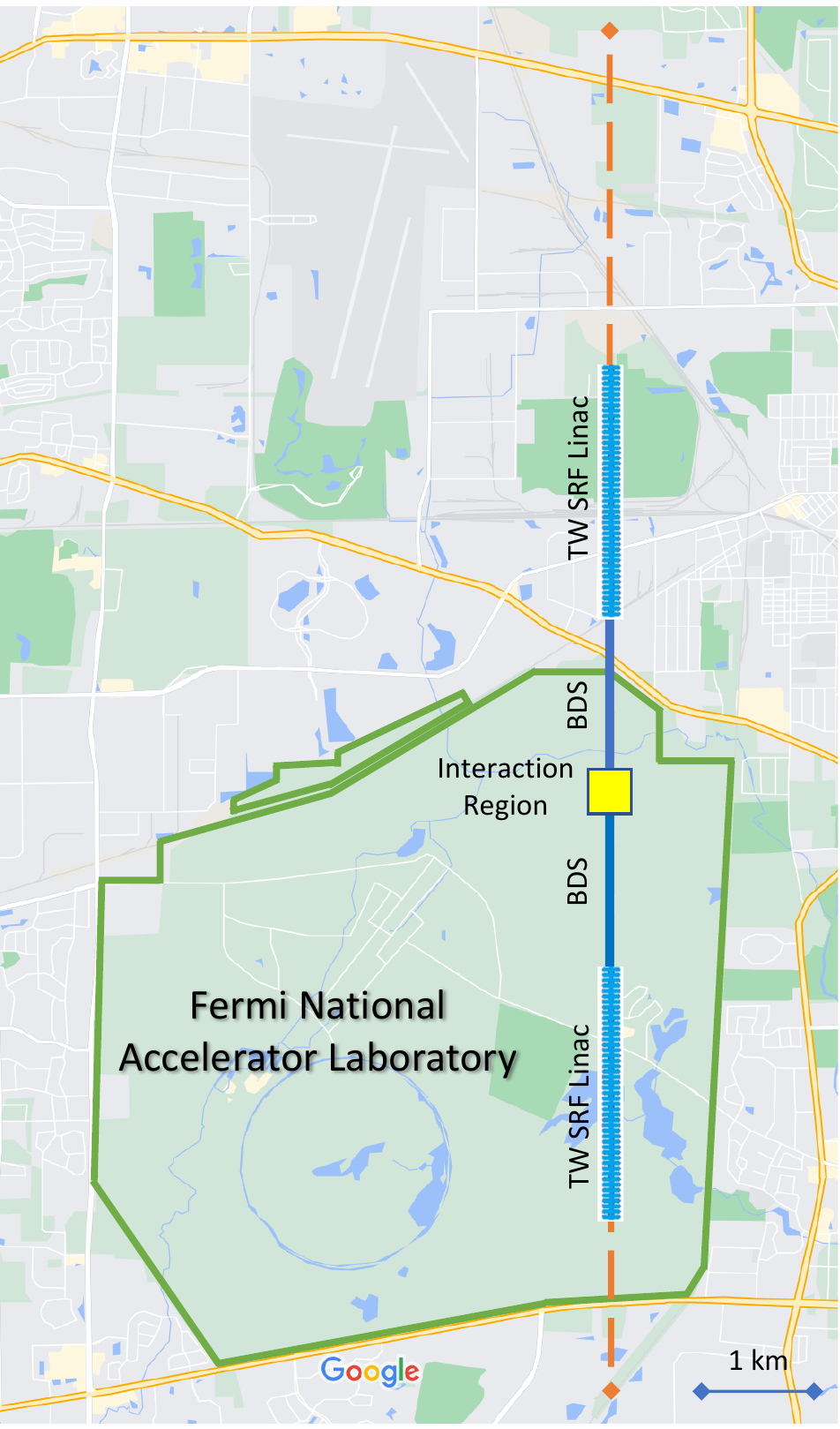}
    \caption{Possible siting of HELEN collider at Fermilab. The TW option is shown. The orange dashed line indicates a 12-km stretch that might be made available for a future linear collider.}
    \label{fig:HELEN_Fermilab}
\end{figure}

\textbf{Possible siting at Fermilab and upgrades.}
As it is discussed in \cite{FutureColliderUS}, there are three possible locations for a linear collider at Fermilab: two 7-km NE--SW orientations fitting diagonally on the site (Figure~3 in \cite{FutureColliderUS}), and a 12-km footprint with N--S orientation extending outside Fermilab boundary, but with the Interaction Region (IR) remaining on site. Option 3 of the HELEN collider will fit into NE--SW locations, while Options 2 and 3 would fit only within the N--S footprint. Figure~\ref{fig:HELEN_Fermilab} show a possible siting of HELEN collider on Fermilab site for the TW option. 

\begin{figure}
    \centering
    \includegraphics[width=0.7\textwidth]{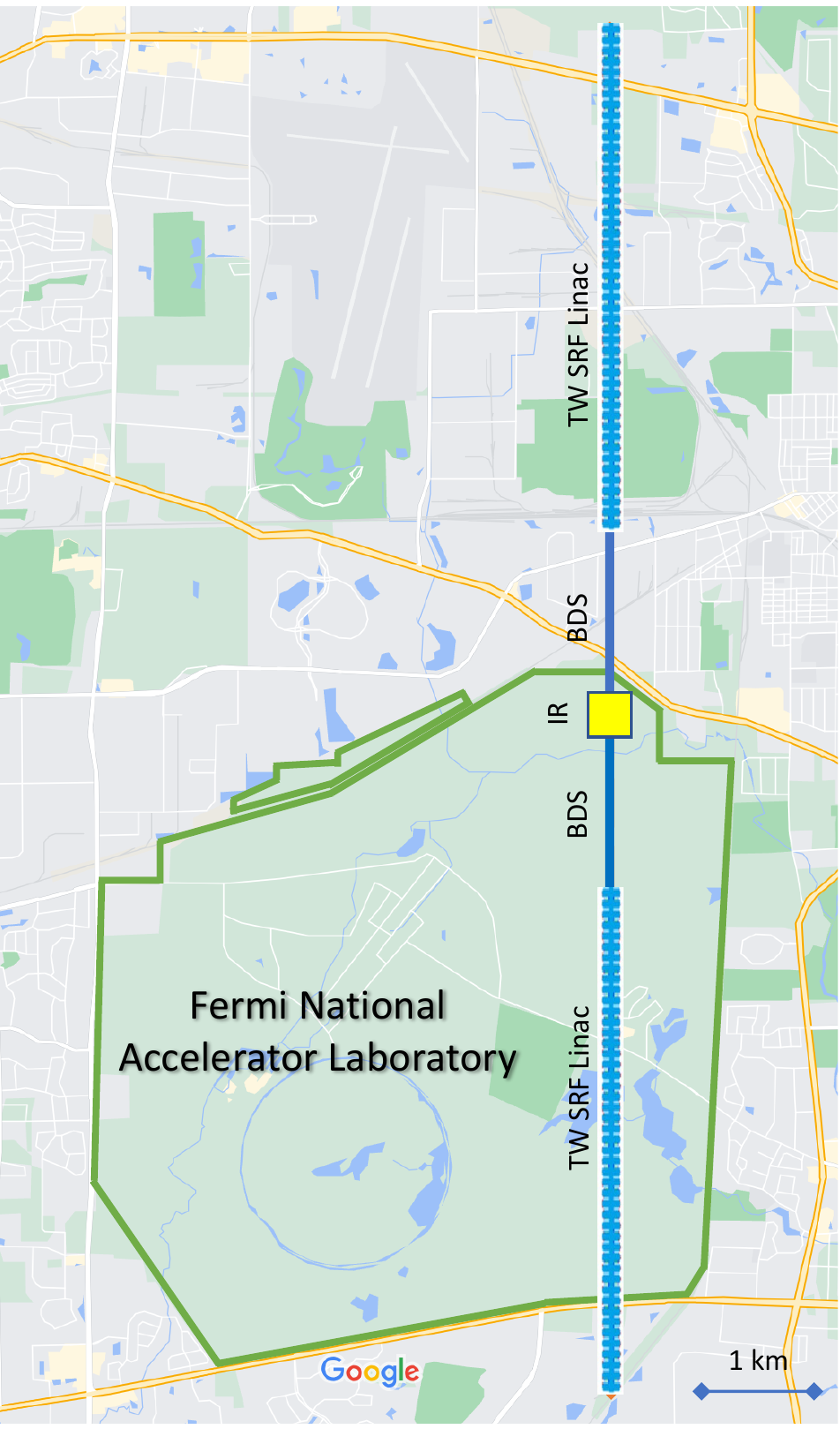}
    \caption{500 GeV HELEN collider at Fermilab.}
    \label{fig:HELEN500_Fermilab}
\end{figure}

As HELEN is similar to ILC in many respects, all proposed ILC luminosity upgrade scenarios (see, e.g., \cite{ILC_Snowmass2021}) are applicable. For possible energy upgrades up to 500~GeV, let us consider only those that could be implemented at Fermilab. To fully utilize the available 12 km, one has to shift the IR further North, closer to the site boundary. For Option 1, the maximum reachable energy is only 350~GeV in this case. Option 2 can be upgraded to 500~GeV (see Figure~\ref{fig:HELEN500_Fermilab}), filling the whole 12 km footprint. Finally, Option 3 can be upgraded to 500~GeV with the 10.8 km collider length.

%% file: Detector.tex
\section{Detector for HELEN Collider \label{sec:Detector}}

A possible detector for the HELEN collider could look very similar to proposed detector concepts for the ILC~\cite{behnke2013international} or CLIC~\cite{Linssen:2012hp,CLIC_YR} and would profit immensely from previous studies and R\&D carried out by those communities. Detailed simulation studies would of course have to be carried out, but by first principles we can assume that the fundamental challenges and requirements would be similar. The physics program determines the detector requirements with some of the key points being high-resolution jet energy reconstruction and di-jet mass resolution to separate W and Z di-jet final states; excellent momentum resolution for charged particles driven by the need to reconstruct the Higgs boson recoiling from leptonically decaying Z bosons; and unprecedented flavor tagging.

In general such a detector would call for a highly-efficient, very low-mass, small-pixel vertexing and tracking system enabling good flavor tagging and heavy and light quark separation, as well as excellent transverse momentum resolution for high-$p_T$ tracks. The latter also requires a large magnetic field on the order of 4 T surrounding the tracking and calorimeter systems. The concept of pulsed power applications, as well as ultra-lightweight cooling and support structures need to be considered in order to keep the overall detector mass in the innermost region at the lowest possible level. In order to enable unprecedented energy resolution, high-granularity electromagnetic and hadronic calorimeters are needed. Detector coverage for electrons and photons needs to extend to very low polar angles to aid in the rejection of large background levels from beamstrahlung. Also for the calorimeters pulsed power readout electronics would be beneficial, taking advantage of the low-duty cycle beam structure of a linear collider, and allowing to forego the need for active cooling. Muon identification would be performed by an instrumented iron return yoke on the outside of the detector. Precision timing on the order of ns might be important for background tagging and pileup rejection, but should be less relevant than for example for CLIC, given the much longer bunch trains, and longer gaps. Radiation hardness is much less a concern than for current LHC experiments, as the expected TID and NIEL levels are going to be orders of magnitude lower. Concerning the data volume and rates, even with continuous, triggerless readout they should be well below current LHC detector rates. 

Fermilab provides a range of detector facilities and technical capabilities that could aid the development of several of these detector components. With SiDet we have a world-class silicon micro- and macro-packaging facility, which has been originally built for the development and construction of the silicon detectors for the Tevatron experiments D0 and CDF. Since then we have built many generations of state-of-the-art semiconductor detectors for collider experiments and astro-particle physics there. We have a large group of ASIC designers in house, who have collaborated on many key chips for HEP experiments. With our scintillator facility we have contributed extruded scintillator to many HEP efforts, as well as performed R\&D into molded and 3D-printed scintillators. The on-site testbeam and irradiation facilities complement the technological expertise.  

%% file: RnD.tex
\section{Accelerator and detector R\&D for HELEN collider \label{sec:RnD}}

\subsection{Accelerator R\&D program objectives}
The traveling wave option seems to be an optimal choice for the HELEN collider, as it provides the best cost and AC power efficiency while delivering gradient high enough to allow energy upgrades up to 500~GeV. Then the major objectives of the accelerator R\&D program should be on advancing the TW SRF technology toward demonstrating its feasibility and culminate in producing a Technical Design Report: 
\begin{itemize}
    \item Demonstrate the feasibility of the TW SRF technology:
    \begin{itemize}
        \item test proof-of-principle 1.3 GHz TW cavity (several cells) and demonstrate accelerating gradient of $\sim 70$~MV/m
        \item adapt an advanced cavity treatment techniques, so that high $Q \sim 10^{10}$ can be achieved at high gradients
        \item design, build and test full-scale prototype cavities; demonstrate performance needed for the HELEN collider
        \item design and build a cryomodule for TW SRF cavities
        \item verify the cryomodule performance without beam on a test stand and with beam at Fermilab's FAST facility
    \end{itemize}
    \item Design and optimize the HELEN linear collider accelerator complex 
    \item Confirm the physics reach and detector performance for the HELEN beam parameters
    \item Publish a Conceptual Design Report as modification of the ILC design in 2--3 years 
    \item Prepare a Technical Design Report after demonstrating the cryomodule performance
\end{itemize}

\subsection{Detector R\&D program objectives}
Given that funding for generic, ``blue sky'' R\&D is scarce, the most practical approach to solving some of the technological limitations for future collider detectors is to identify appropriate intermediate projects (e.g. future LHCb and ALICE upgrades, EIC, etc.) that could serve as stepping stones, or small- to mid-scale technology demonstrators. For example for the inner vertexing detector, R\&D on various technologies is already going on, with advances in CMOS detectors taking most of the recent focus. Especially recent developments by Mu3e~\cite{Mu3e} 
and ALICE~\cite{ALICE} into ultra-lightweight support structures made of Kapton, or wafer-sized, bent sensors, are of strong relevance here. Furthermore, R\&D into novel, room-temperature refrigerants, air cooling, silicon micro-channel cooling directly etched into the silicon sensor's backside, or carbon-based cooling pipes lay out promising R\&D directions. For the highly-granular calorimeters, which are essential to support necessary particle flow algorithms, several technologies are being investigated. Many of which have been demonstrated over many years within the CALICE collaboration~\cite{CALICE}. One specific application is currently being built within the context of the HL-LHC CMS Detector Upgrade. The main challenges here are the overall size, complexity and cost of the detector. Some of the technology choices available are silicon as sensitive material, liquid argon calorimetry, or dual readout calorimeters containing fibers or crystals. Each of these has advantages and disadvantages related to energy resolution, containment and granularity. One particular recent interest that has emerged is the addition of fast timing information within the calorimeter, where 20--50 ps timing resolution per hit would enable 5D calorimetry and thus tracking inside showers. R\&D needs to be performed on LGAD sensors to optimize sensors thickness with regard to signal size and timing resolution.